\documentclass{aa}  

\usepackage{enumitem}
\usepackage{amsmath}
\usepackage{xcolor}

\def\gmrt {\emph{GMRT}}

\def\lax{\lesssim}
\def\gax{\gtrsim}

\usepackage{txfonts}
\usepackage{hyperref}
\begin{document}

   \title{The Multi-Epoch Jet Outbursts in Abell~496: synchrotron ageing and buoyant X-ray cavities draped by warm gas filaments}

    \titlerunning{The Multi-Epoch Jet Outbursts in Abell~496}
   \author{F. Ubertosi
          \inst{1,2}\fnmsep\thanks{Contact: \texttt{francesco.ubertosi2@unibo.it}}
          \and
          S. Giacintucci\inst{3}
          \and 
          T. Clarke\inst{3}
          \and
          M. Markevitch\inst{4}
          \and
          T. Venturi\inst{5}
          \and
          E. O'Sullivan\inst{6}
          \and
          M. Gitti\inst{1,5}
          }

   \institute{Dipartimento di Fisica e Astronomia, Università di Bologna, via Gobetti 93/2, I-40129 Bologna, Italy,
         \and
             Istituto Nazionale di Astrofisica - Osservatorio di Astrofisica e Scienza dello Spazio (OAS), via Gobetti 101, I-40129 Bologna, Italy,
        \and
            Naval Research Laboratory, 4555 Overlook Avenue SW, Code 7213, Washington, DC 20375, USA,
        \and
            NASA/Goddard Space Flight Center, Greenbelt, MD 20771, USA,
        \and 
            Istituto Nazionale di Astrofisica - Istituto di Radioastronomia (IRA), via Gobetti 101, I-40129 Bologna, Italy,
        \and
            Center for Astrophysics $|$ Harvard \& Smithsonian, 60 Garden Street, Cambridge, MA 02138, USA.
             }

   \date{Received: August 2, 2024; Accepted: September 20, 2024}

 
  \abstract
   {}
   {The galaxy cluster Abell~496 has been extensively studied in the past for the clear sloshing motion of the hot intracluster medium (ICM) on large scales, but the interplay between the central radio galaxy and the surrounding cluster atmosphere is mostly unexplored. We present a dedicated radio, X-ray, and optical study of Abell~496 aimed at investigating this connection.
   }
   {We use deep radio images obtained with the Giant Metrewave Radio Telescope (GMRT) at 150, 330 and 617 MHz, Very Large Array (VLA) at 1.4 and 4.8 GHz, and VLA Low Band Ionosphere and Transient Experiment (VLITE) at 340 MHz, with angular resolutions ranging from $0.^{\prime\prime}5$ to $25^{\prime\prime}$. Additionally, we use archival {\it Chandra} and Very Large Telescope (VLT) MUSE observations.}
   {The radio images reveal three distinct periods of jet activity: an ongoing episode on sub-kpc scales with an inverted radio spectrum; an older episode that produced lobes on scales $\sim20$~kpc which now have a steep spectral index ($\alpha=2.0\pm0.1$);  and an oldest episode that produced lobes on scales of $\sim50-100$~kpc with an ultra-steep spectrum ($\alpha=2.7\pm0.2$). Archival {\it Chandra} X-ray observations show that the older and oldest episodes have excavated two generations of cavities in the hot gas of the cluster. The outermost X-ray cavity has a clear mushroom-head shape, likely caused by its buoyant rise in the cluster's potential. Cooling of the hot gas is ongoing in the innermost 20~kpc, where H$\alpha$-bright warm filaments are visible in VLT-MUSE data. The H$\alpha$-filaments are stretched towards the mushroom-head cavity, which may have stimulated ICM cooling in its wake. We conclude by commenting on the non-detection of a radio mini-halo in this vigorously sloshing, but low-mass, galaxy cluster.}
   {}


   \maketitle
%
\section{Introduction}
Radio galaxies represent clear manifestations of jet launching from active galactic nuclei (AGN). The lobes of these objects, extending on either side of the central engine, are very informative on how AGN interact with and influence their surrounding environment (e.g., \citealt{2020NewAR..8801539H,2023Galax..11...73B}). This is especially relevant for radio galaxies at the center of galaxy clusters with cool cores. Cool core clusters, characterized by a central region of lower gas temperature compared to the surrounding intracluster medium (ICM), present a unique environment for radio lobe evolution. The expanding radio lobes often push aside the ICM, creating depressions in the X-ray emitting gas (e.g., \citealt{2007ARA&A..45..117M,2012NJPh...14e5023M}). 

The active phase of a super massive black hole (SMBH) is not continuous. Periods of intense jet activity can be followed by quiescent phases, leaving behind the remnants of past outbursts (e.g., \citealt{2020NewAR..8801539H}). After detaching from the jets, the radio lobes age and further expand in the ICM. Radio-filled cavities have a lower density than the surrounding ICM, leading to a buoyant rise of these structures towards the cool core outskirts ($\sim$100~kpc; e.g., \citealt{2001ApJ...554..261C}). As these bubbles rise, they can entrain the surrounding ICM in their wake, stimulating cooling instabilities in the hot gas and condensation to cooler gas phases (e.g., \citealt{2015ApJ...799L...1V}). The evolution of bubbles at later times is more uncertain: theoretical models predict that efficient shear instabilities and mixing should lead to bubbles losing their integrity and eventually disrupting, unless viscosity or magnetic field stabilize them against instabilities (e.g., \citealt{2005MNRAS.357..242R,2006MNRAS.371.1025S,2015ApJ...802..118B}). 

Addressing these issues with observations requires a fortunate combination of sensitivity, spatial resolution, and frequency coverage, because (a) the oldest radio lobes are inherently fainter due to energy losses in the relativistic electrons, the magnitude of which can only be determined by employing multi-frequency radio observations; (b) the corresponding X-ray cavities reside in the outskirts of, or beyond, the cool core, thus their contrast in X-ray images is relatively lower; (c) resolving the features caused by instabilities requires high spatial resolution. So far, the stability of buoyant bubbles has been studied only for the radio lobes/bubbles in Virgo and Perseus (e.g., \citealt{2001ApJ...554..261C,2005MNRAS.359..493K,2013MNRAS.436.1721R}), which are respectively the closest and X-ray brightest clusters. 

In this work, we address the above topics in Abell 496 (hereafter A496), a nearby cool core galaxy cluster ($z = 0.0329$, cooling radius $\sim$70~kpc) extensively studied in the X-ray band for the clear bulk motions ({\em sloshing}) of its ICM on large scales \citep{2003ApJ...583L..13D,2006PASJ...58..703T,2007ApJ...671..181D,2012MNRAS.420.3632R,2014A&A...570A.117G}. One of the most interesting feature of this cluster is the irregular and boxy shape of its cold fronts (surface brightness edges caused by the bulk motions of the gas), which suggests the presence of Kelvin-Helmholtz instabilities in the ICM at the interface of the edges \citep{2012MNRAS.420.3632R,2014A&A...570A.117G}. However, the properties of the radio galaxy at the center of this cluster were so far neglected. The available information are mainly from the sample studies of \citet{2015MNRAS.453.1201H,2015MNRAS.453.1223H}, who report the presence of radio emission with an ultra-steep radio spectrum at sub-GHz frequencies (indicative of particles aging), and of radio emission with a flat spectrum above 1 GHz (indicative of renewed SMBH activity). In this paper, we leverage multi-frequency radio observations of A496, as well as X-ray and H$\alpha$ data, to study the central radio source of this cluster and its interaction with the hot gas. 

We summarize the general properties of A496 in Table~\ref{tab:a496}. We adopt $\Lambda$CDM cosmology with H$_0$=70~km~s$^{-1}$~Mpc$^{-1}$, $\Omega_m=0.3$ and $\Omega_{\Lambda}=0.7$. At the redshift of A496 ($z=0.0329$), $1^{\prime\prime}$ corresponds to 0.656 kpc. The radio spectral index $\alpha$ is defined according to $S_{\nu} \propto \nu^{-\alpha}$, where $S_{\nu}$ is the flux density at the frequency $\nu$.


\begin{table}[t]\label{tab:a496}
\caption{Properties of the galaxy cluster A496}
\begin{center}
\begin{tabular}{lc}
\hline\noalign{\smallskip}
\hline\noalign{\smallskip}
& Value \\
\hline\noalign{\smallskip}
R.A.$_{\rm J2000}$ (h m s)  &  04 33 38.4  \\
Decl.$_{\rm J2000}$ ($^{\circ}$ $^{\prime}$ $^{\prime\prime}$) & $-13$ 15 33 \\
\phantom{0}$z$ &  0.0329 \\
\phantom{0}$D_L$ (Mpc) & 144.5  \\
\phantom{0}Linear scale (kpc/$^{\prime \prime}$) &  0.656    \\ 
$M_{\rm 500}$ ($10^{14}$ $M_{\odot}$) & $2.7$   \\
\noalign{\smallskip}
\hline\noalign{\smallskip}
\end{tabular}
\end{center}
\label{tab:sources}
\tablefoot{ The total mass within $R_{\rm 500}$ is from 
\cite{2014A&A...571A..29P}, where $R_{500}$ is the radius 
within which the cluster mean total density is 500 times the critical density at the cluster redshift.}
\end{table}



\begin{table*}[ht!]
    \renewcommand{\arraystretch}{1.2}
            \caption{Summary of the radio observations}\label{tab:obs2}
                 \centering
    \begin{tabular}{c|c|c|c|c|c|c|c}

    \hline
         Radio & Project & Target & $\nu$ & $\Delta\nu$ & Observation & Duration & Primary \\ 
    telescope &    code & name &(MHz)  & (MHz)  &  date  &   (min) & calibrators \\
\hline
GMRT\phantom{0}    &  $19_{-}043$\phantom{0}& R14D23 &  150  & 16 (14) & 2011 Jan 24 &  15 & 3C48, 3C286    \\ 
GMRT\phantom{0}    &  $19_{-}043$\phantom{0}& R14D24 &  150  & 16 (14) & 2011 Jan 24 &  15 & 3C48, 3C286     \\ 
GMRT\phantom{0}    &  $19_{-}043$\phantom{0}& R14D25 & 150  & 16 (14) & 2011 Jan 24 &  15 &  3C48, 3C286     \\ 
GMRT\phantom{0}    &  $17_{-}073$\phantom{0}& A496   &   330  & 16 (11) & 2010 Jan 15 & 124 &  3C48, 3C286  \\ 
GMRT\phantom{0}    &  05DAG01\phantom{0}   &  A496   &   617  & 32 (30) & 2004 Mar 22 & 78  &  3C48         \\ 
VLITE--A\phantom{0}   &  20B-138$^a$     &  A496  & 340  & 64 (38) & 2021 Feb 15   & 102  & 3C48, 3C147, 3C286  \\
VLA--B\phantom{0}  & AC540\phantom{0}      & 0434$-$131  &  1365/1465 & 25/25      & 2000 Jan 04  & 147 &  3C138, 3C147 \\
VLA--B\phantom{0}  & AC540\phantom{0}      & 0434$-$131  &  1365/1465 & 25/25      & 2000 Jan 07  & 147 & 3C138, 3C147 \\
VLA--A\phantom{0}  & AC572\phantom{0}       & 0431$-$132  &   1465/1515 & 25/25      & 2000 Dec 13 & 113 &  3C286 \\  
VLA--A\phantom{0}  & A074\phantom{0}        &    A496     &   4835/4885 & 50/50      & 1987 Sep 22 &  9  &  3C138 \\ 
VLA--B\phantom{0}  & AB398\phantom{0}       &  A496       &   4823/4873 & 25/25      & 1986 Sep 07 &   8 &  3C286 \\
VLA--C\phantom{0}  & AC572\phantom{0}       & 0431$-$1321 &   4535/4885 & 50/50      & 2001 Jun 21 &  90 &  3C48 \\ 
VLA--D\phantom{0}  & AS220\phantom{0}       &  A496       &  4835/4885 & 50/50      & 1985 Dec 18 &  20 & 3C285 \\
\hline
    \end{tabular}
\tablefoot{Column 1: radio telescope (the array configuration is reported for the VLA and VLITE). Column 2: project code. Column 3: target name. Columns 4--5: observing frequency and bandwidth. For the VLA observations, we report the frequency and bandwidth of the two were Intermediate Frequency channels. For the GMRT and VLITE datasets, 
the usable bandwidth after data flagging is reported in parenthesis. Columns 6--7: observing date and time. Column 8: calibration sources used to set the flux density scale (see text for details).
$a$: Primary VLA project during which VLITE was operational.}
 
\end{table*}

\begin{table}[]
\renewcommand{\arraystretch}{1.2}
    \centering
    \setlength{\tabcolsep}{3pt}
    \caption{Summary of the radio images}\label{tab:images}
    \begin{tabular}{c|c|c|c|c}
\hline
 {Radio} & 
 {$\nu$} & 
 {FWHM}  & 
 {rms}   & 
 {Notes} \\
 {telescope} &
  {(MHz)}     &   
  {($^{\prime \prime} \times^{\prime \prime}$)} &  
  {(mJy/beam)} &  
 \\
 \hline
 GMRT\phantom{0}  & 150  & $24\times18$ & 10  & Fig.~3b \\
GMRT\phantom{0}  & 330  & $14\times9$  & 0.5  & Fig.~3a \\
GMRT\phantom{0}  & 617  & $6\times5$   & 0.35 & Fig.~2b \\
VLITE--A\phantom{0}   &  340  & $7\times4$  & 0.66 &  Fig.~2a \\
VLA--A\phantom{0}  & 1490  & $2\times1$   & 0.05 & Fig.~1\\
VLA--B\phantom{0}  & 1415  & $5\times5$   & 0.1 & Fig.~2c \\
VLA--A\phantom{0}  & 4860 & $0.6\times0.4$ & 0.11 & not shown \\
VLA--B\phantom{0}  & 4848 & $1.7\times1.1$   & 0.15 & not shown\\
VLA--C\phantom{0}  & 4710 & $4\times4$   & 0.03 & Fig.~2d \\
\hline
    \end{tabular}
    \tablefoot{Column 1: radio telescope (the array configuration is reported 
for the VLA and VLITE). Columns 2--3: central frequency, full width half maximum (FWHM) of 
the radio beam, and rms noise level ($1\sigma$). All images were made using a Briggs 
robust parameter of 0.}
    
\end{table}

%
%
%
\begin{figure}
\centering
\includegraphics[trim={0.6cm 1cm 2cm 0.5cm},clip,width=0.9\linewidth]{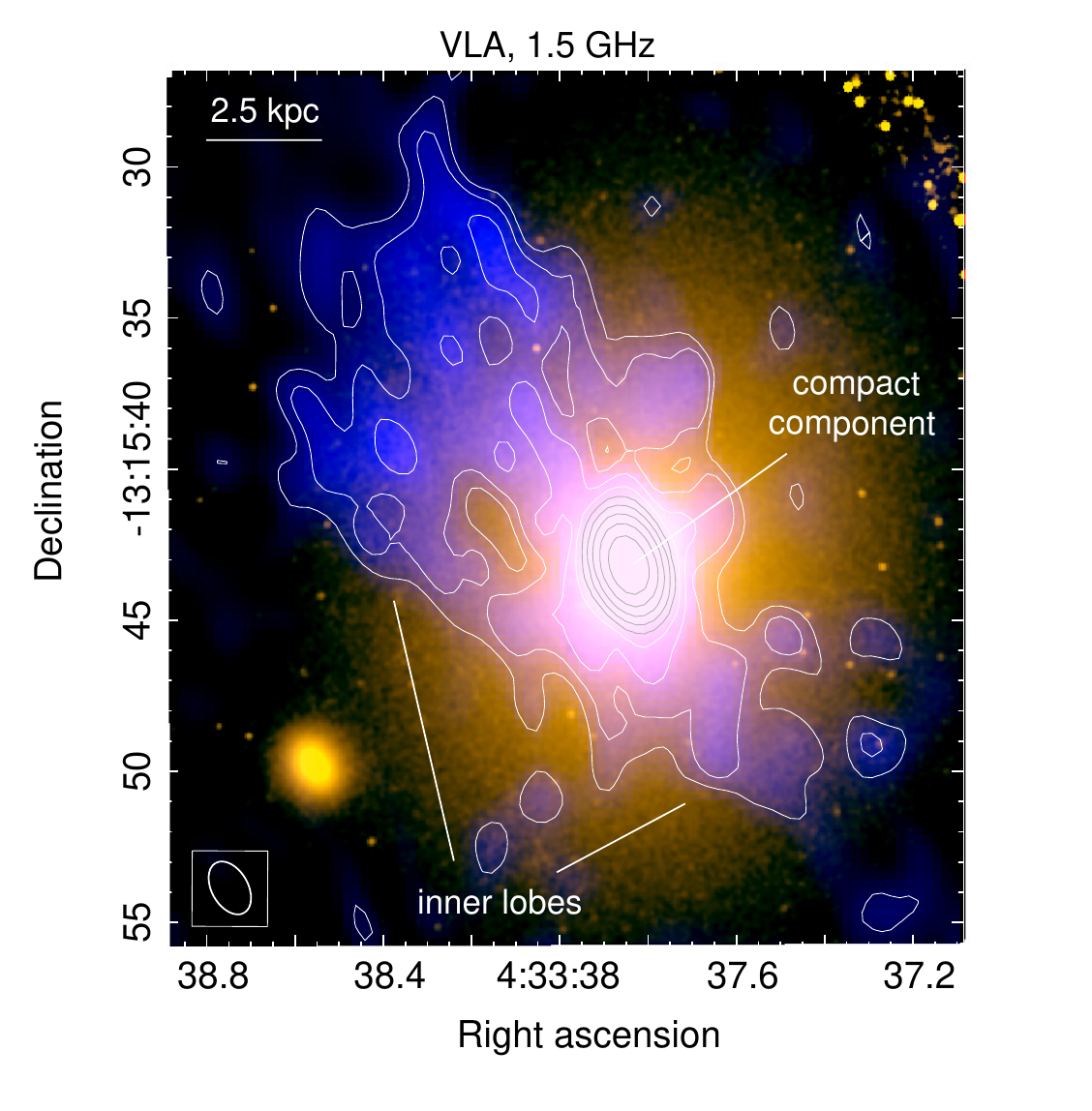}
\caption{VLA A--configuration image at 1.5 GHz (blue and contours) of the central 
radio source in A496, overlaid on the optical {\em HST} WFPC2 image 
of the central galaxy (orange). The restoring beam of the radio image is $2^{\prime\prime}\times1^{\prime\prime}$, 
in p.a. $17^{\circ}$ (boxed white ellipse in the bottom left corner) and the noise level is 
$1\sigma=50$ $\mu$Jy beam$^{-1}$. Contours are spaced by a factor of 2 starting
from $+3\sigma$.}
\label{fig:hst}
\end{figure}

\section{Radio observations}\label{sec:obs}

We analyzed pointed observations of A496 retrieved from the Giant Metrewave Radio Telescope (GMRT) and Very Large Array (VLA) archives, 
covering a 
frequency interval from 330 MHz to 4.9 GHz and with an angular resolution 
ranging from $\sim 0.^{\prime\prime}$5 to $\sim 20^{\prime\prime}$. Our 
analysis also includes data at 340 MHz from the VLA Low Band Ionosphere 
and Transient Experiment \citep[VLITE,][]{2016SPIE.9906E..5BC}, 
a commensal system that operates on the VLA. Finally, we re-analyzed GMRT 
data at 150 MHz from 3 fields of the TIFR GMRT Sky Survey 
(TGSS\footnote{\url{http://tgss.ncra.tifr.res.in/}}) that contain A496 within
their field of view. Table~\ref{tab:obs2} summarized the details on all 
radio observations processed in this paper.

\subsection{GMRT data}

We retrieved observations at 150 MHz, 330 MHz and 617 MHz from the GMRT archive. 
All data were collected in spectral-line mode using the hardware back--end.
A total observing bandwidth of 16 MHz was used at 150 MHz and 330 MHz. 
Two 16 MHz bands (upper and lower side bands, USB/LSB) were instead used 
at 617 MHz, for a total bandwidth of 32 MHz. Below, we provide a description of 
the data reduction at each of these frequencies.

\subsubsection{Observations at 150 MHz and 330 MHz} 

We reduced the data using the National Radio Astronomy Observatory (NRAO) Astronomical 
Image Processing System \citep[AIPS\footnote{\url{http://www.aips.nrao.edu}.},][]{2003ASSL..285..109G}.
We used RFLAG to excise visibilities affected by radio frequency interference (RFI),
followed by manual flagging to remove residual bad data. Gain and bandpass calibrations 
were applied using the primary calibrators (Table~\ref{tab:obs2}). Their flux density 
at 150 MHz and 330 MHz was set based on the \cite{2012MNRAS.423L..30S} scale. 
Phase calibrators, observed several times during the observation, were used to calibrate 
the data in phase. A number of phase self-calibration cycles, followed by a final 
self-calibration step in amplitude, were applied to the target visibilities. 
Non-coplanar effects were taken into account using wide-field imaging at all 
frequencies, decomposing the primary beam area (half-power beamwidth $\sim 186^{\prime}$ at 150 MHz and 
$\sim 81^{\prime}$ at 330 MHz) into smaller facets. Additional facets were placed on 
outlier bright sources out to a radius of $10^{\circ}$ from the center. 
Small clean masks were placed by hand around all sources in the imaged field. 
Finally, to improve the dynamic range of our images, we used PEELR 
to {\em peel} \citep[e.g.,][]{2004SPIE.5489..817N} a bright radio source (PKS0431--133),
located $\sim 10^{\prime}$ from the A496 center, whose strong sidelobe pattern 
affects the cluster region at all frequencies. At 150 MHz, the 3 TGSS fields 
were calibrated and imaged separately. The final images were restored with a 
common beam of $24^{\prime\prime}\times18^{\prime\prime}$, corrected for the 
GMRT primary-beam attenuation, and finally combined together in the image plane. 

The primary beam correction was done with PBCOR using the appropriate 
primary-beam shape parameters at each frequency\footnote{\url{http://www.ncra.tifr.res.in:8081/~ngk/primarybeam/beam.html}.}.
Table~\ref{tab:images} summarizes the restoring beams and root mean square (rms) noise levels ($1\sigma$) 
of our final GMRT images, obtained setting the Briggs robustness 
weighting \citep{1995PhDT.......238B} to 0 during the clean in IMAGR. Residual amplitude errors are estimated to be within $15\%$ at 150 MHz and $10\%$ at 327 MHz 
\citep{2004ApJ...612..974C}.

\subsubsection{Observations at 617 MHz} 

We used the Source Peeling and Atmospheric Modeling \citep[SPAM;][]{2009A&A...501.1185I}
pipeline\footnote{\url{http://www.intema.nl/doku.php?id=huibintemaspampipeline}.} to process the
archival GMRT observation at 617 MHz (Table~\ref{tab:obs2}). 
We adopted a standard calibration scheme that consists 
of bandpass and complex gain calibration and direction-independent self-calibration, 
followed by direction-dependent self-calibration. The flux density scale was set using 3C\,48 and 
\cite{2012MNRAS.423L..30S}. The USB and LSB datasets were processed individually. 
The final self-calibrated data sets were then converted into measurement sets using the 
Common Astronomy Software Applications \citep[CASA\footnote{\url{https://casa.nrao.edu}.}, 
version 5.1][]{2022PASP..134k4501C} and finally imaged together using the joint and 
multi-scale deconvolution options in WSClean \citep{2014MNRAS.444..606O, 2017MNRAS.471..301O}
with a Briggs robust parameters of $0$. Correction for the \gmrt\ primary-beam response 
was applied using the AIPS task PBCOR. 
Our final image at 617 MHz has a beam of $6^{\prime\prime}\times5^{\prime\prime}$ 
and a noise of 0.35 mJy/beam (Table~\ref{tab:images}). The systematic 
amplitude uncertainty is estimated to be within 10$\%$ \citep{2004ApJ...612..974C}.

\subsection{VLA data}

We reprocessed 7 archival VLA multi-configuration and multi-frequency observations 
of A496 (Table~\ref{tab:obs2}) using AIPS and following a standard procedure of gain 
and bandpass calibration. The flux density scale was set using SETJY and the 
\cite{2017ApJS..230....7P} coefficients for the primary calibrator sources 
listed in Table~\ref{tab:obs2}. 
The data were calibrated in phase using phase calibration sources observed during the 
observations. Several loops of imaging and phase-only self-calibration were applied 
to reduce the effects of residual phase and amplitude errors in the data. 
During the imaging process, we used small clean masks around the sources. 
All images were made using a Briggs robust parameter of 0. 
The two B-configuration images at 1.4 GHz were first restored with a common $4^{\prime\prime}$ 
circular beam and then combined together into a final image with a 
noise of $1\sigma=100$ $\mu$Jy beam$^{-1}$. Correction for the VLA primary beam 
attenuation was applied to all images. Table~\ref{tab:images} summarizes restoring beams and rms values. Residual amplitude errors are within $5\%$ at all frequencies.

\subsection{VLITE data}\label{sec:vlite}

VLITE is a commensal, low-frequency system on the VLA \cite{2016SPIE.9906E..5BC}. 
It operates in parallel with nearly all regular observing programs above 1 GHz and provides 
real-time correlation of the signal from a subset of VLA antennas (up to 18) 
using the low-band receiver system \citep{2011ursi.confE...5C} and a 
dedicated DiFX-based software \citep{2007PASP..119..318D}. 
The VLITE system processes 64 MHz of bandwidth centered on 352 MHz, however,
due to strong RFI in the upper portion of the band, the usable frequency range 
is limited to an RFI-free band of $\sim 40$ MHz, centered on 338 MHz.

On 2021 February 15, VLITE recorded data with 17 antennas during a VLA 
observation at the primary frequency of 1.5 GHz of A496 (project 20B-138; 
Table~\ref{tab:obs2}), when the VLA was in A configuration. The VLITE data 
were processed using a dedicated calibration and imaging pipeline, 
which is based on a combination of Obit \citep{2008PASP..120..439C} 
and AIPS data reduction packages. The pipeline uses standard automated 
tasks for the removal of RFI, followed by delay, complex gain, and 
bandpass calibration \citep[for details see][]{2016ApJ...832...60P}. 
The flux density scale is set using \cite{2017ApJS..230....7P} and
residual amplitude errors are estimated to be within $15\%$ 
\citep{2016SPIE.9906E..5BC}. The calibrated data are imaged 
using wide-frequency imaging algorithms in Obit (task MFImage), 
by covering the full primary beam with facets and placing outlier facets 
on bright sources out to a radius of $20^{\circ}$. Clean masks are placed 
on the sources during the imaging process to reduce the effects of CLEAN bias. 
The pipeline runs two imaging and phase self-calibration cycles before a 
final image is created. 

For A496, the pipeline produced an image with a beam of $7^{\prime\prime}\times4^{\prime\prime}$ 
and $1.3$ mJy/beam rms. To increase the image sensitivity, 
we re-imaged the pipeline-calibrated data with WSClean 
using a higher number of cleaning iterations than in the 
automated pipeline and a cleaning mask generated from the 
pipeline image. Our final VLITE image has a resolution of 7$^{\prime\prime}\times4^{\prime\prime}$ 
(for ROBUST$=0$) and an improved sensitivity of 
$1\sigma=0.66$ mJy/beam (Table~\ref{tab:images}).

\begin{figure*}[ht!]
\centering
\includegraphics[width=8.5cm]{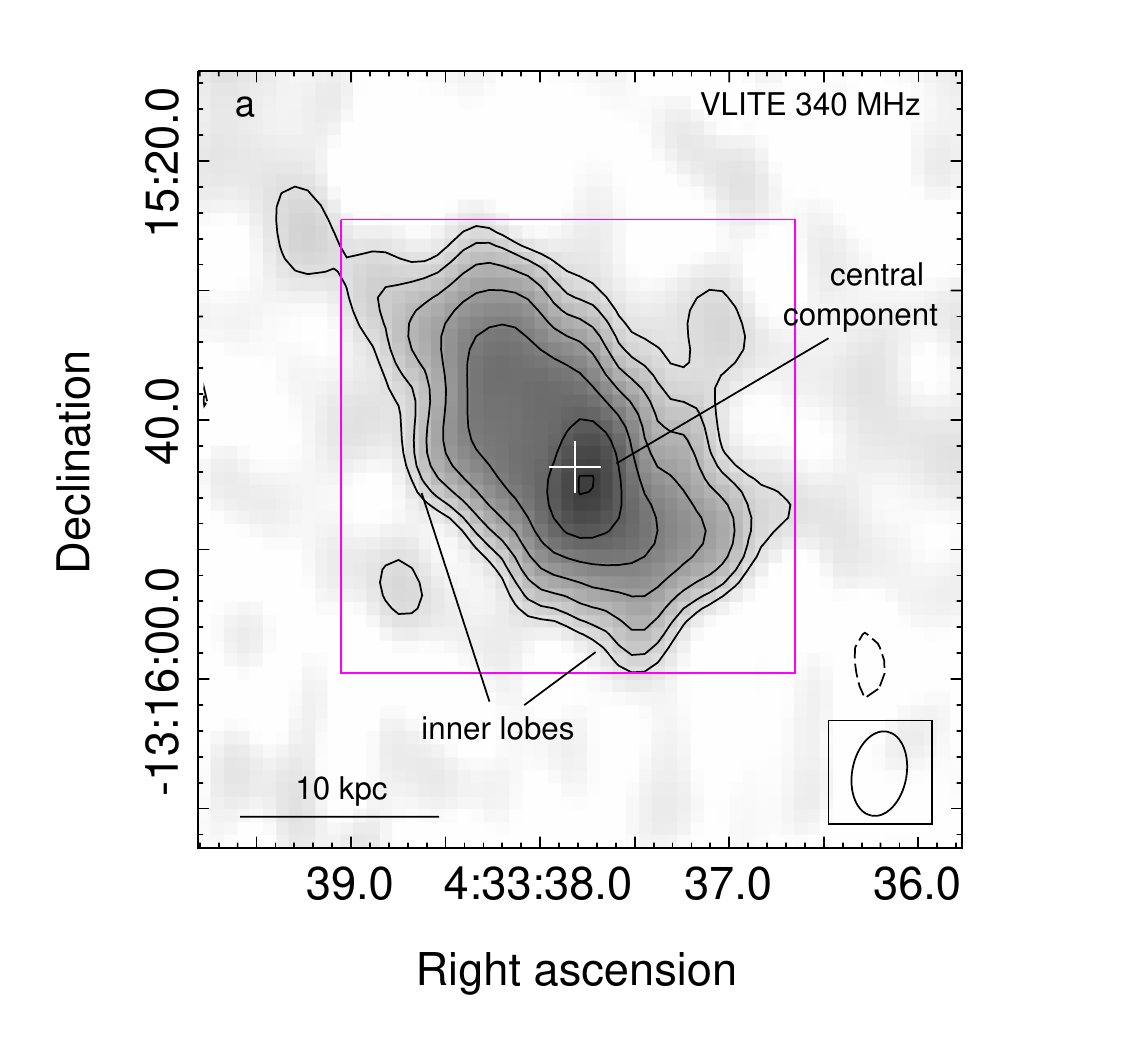}
\includegraphics[width=8.5cm]{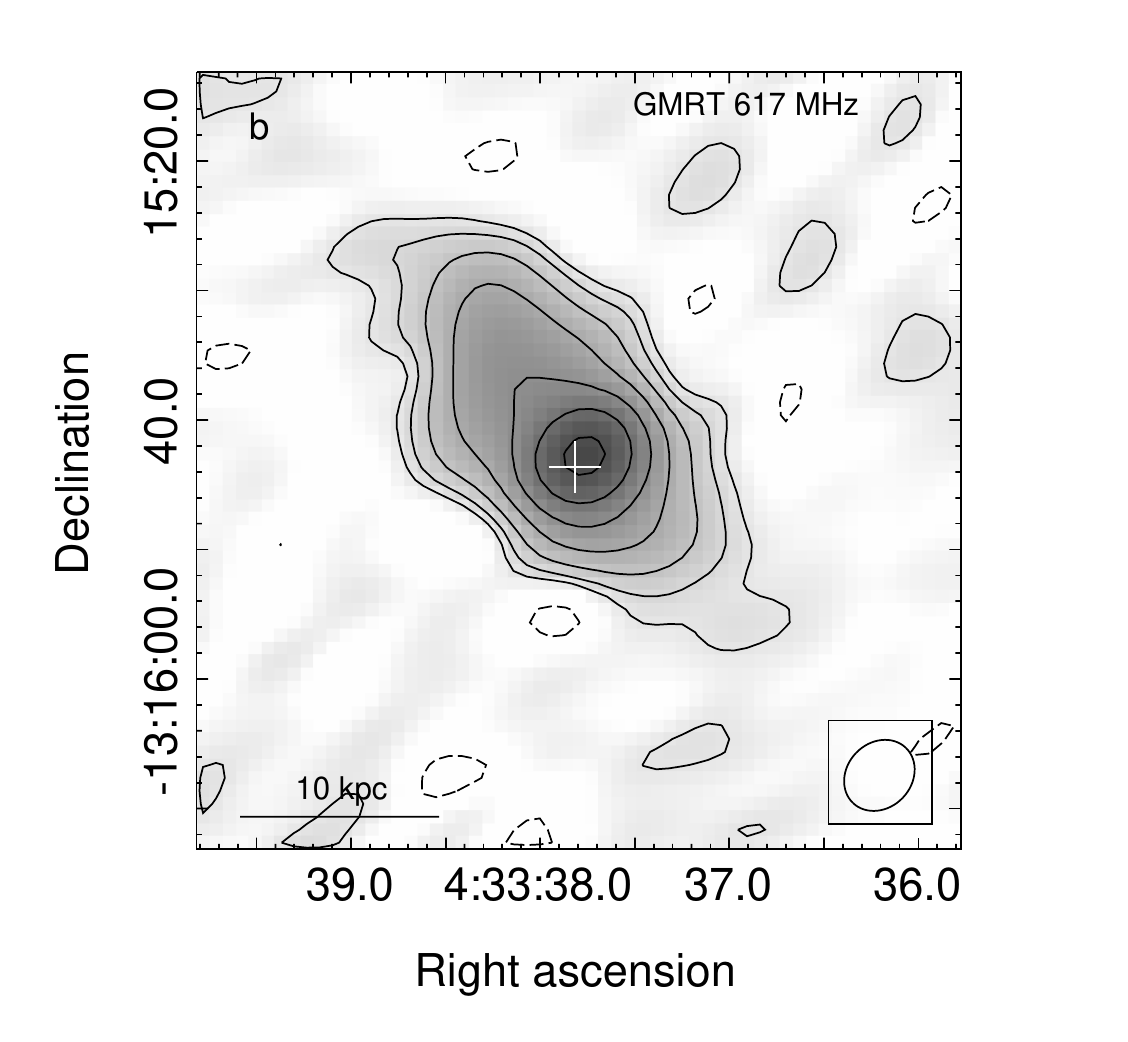}
\includegraphics[width=8.5cm]{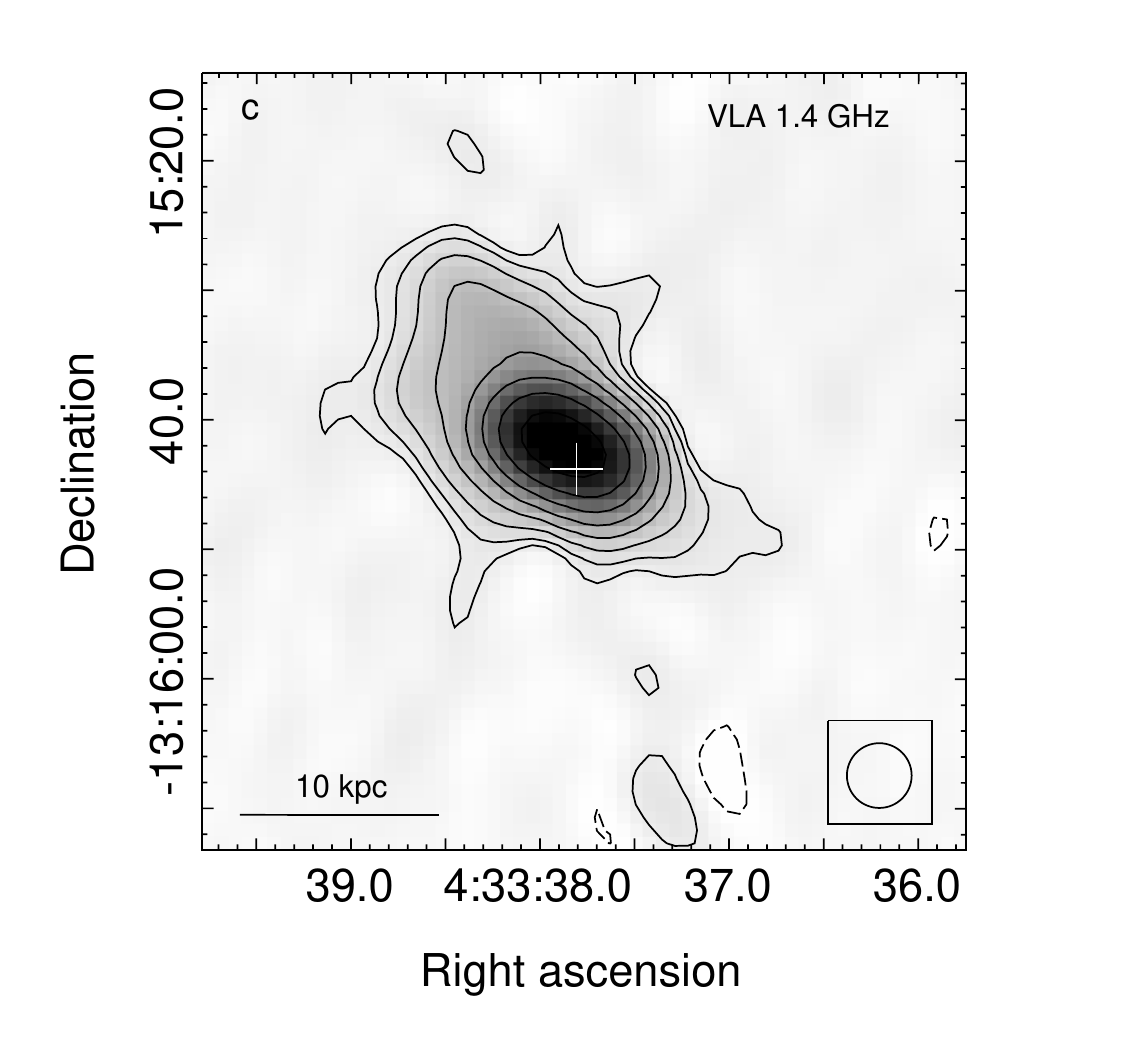}
\includegraphics[width=8.5cm]{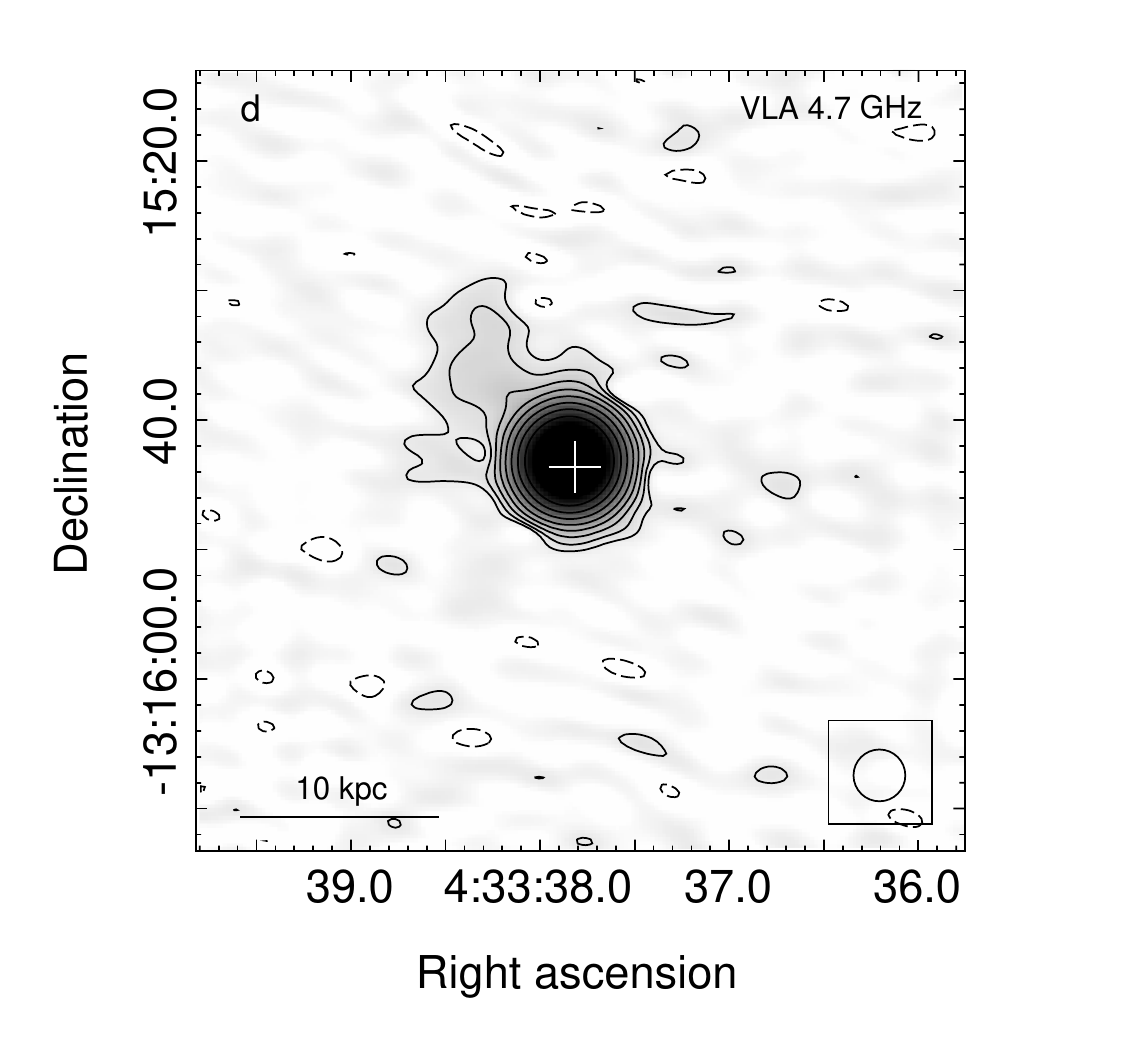}
\smallskip
\caption{a: VLITE 340 MHz image. The beam is $7^{\prime\prime}\times4^{\prime\prime}$, 
in p.a. $-2^{\circ}$ and noise is $1\sigma=0.66$ mJy/beam. 
b: GMRT 617 MHz image. The beam is $6^{\prime\prime}\times5^{\prime\prime}$, in p.a. $-54^{\circ}$ and $1\sigma=0.47$ mJy/beam. c: VLA--B configuration combined image at 1.4 GHz. The image 
has been restored with a $5^{\prime\prime}$ circular beam. The noise is $1\sigma=0.1$ mJy/beam. .
d: VLA--C configuration image at 4.7 GHz. The image has been restored with a $4^{\prime\prime}$ circular beam.
The noise is $1\sigma=0.03$ mJy/beam. In all panels, contours are spaced by a factor of 2 starting from 
$+3\sigma$. Contours at $-3\sigma$ mJy/beam are shown as dashed. The boxed ellipse/circle shows the beam size. The white cross marks the optical peak (Fig.~\ref{fig:hst}). In panel a, the magenta box is as in Fig.~\ref{fig:low}}. 
\label{fig:617}
\end{figure*}

\begin{figure*}[ht!]
\centering
\includegraphics[width=8.5cm]{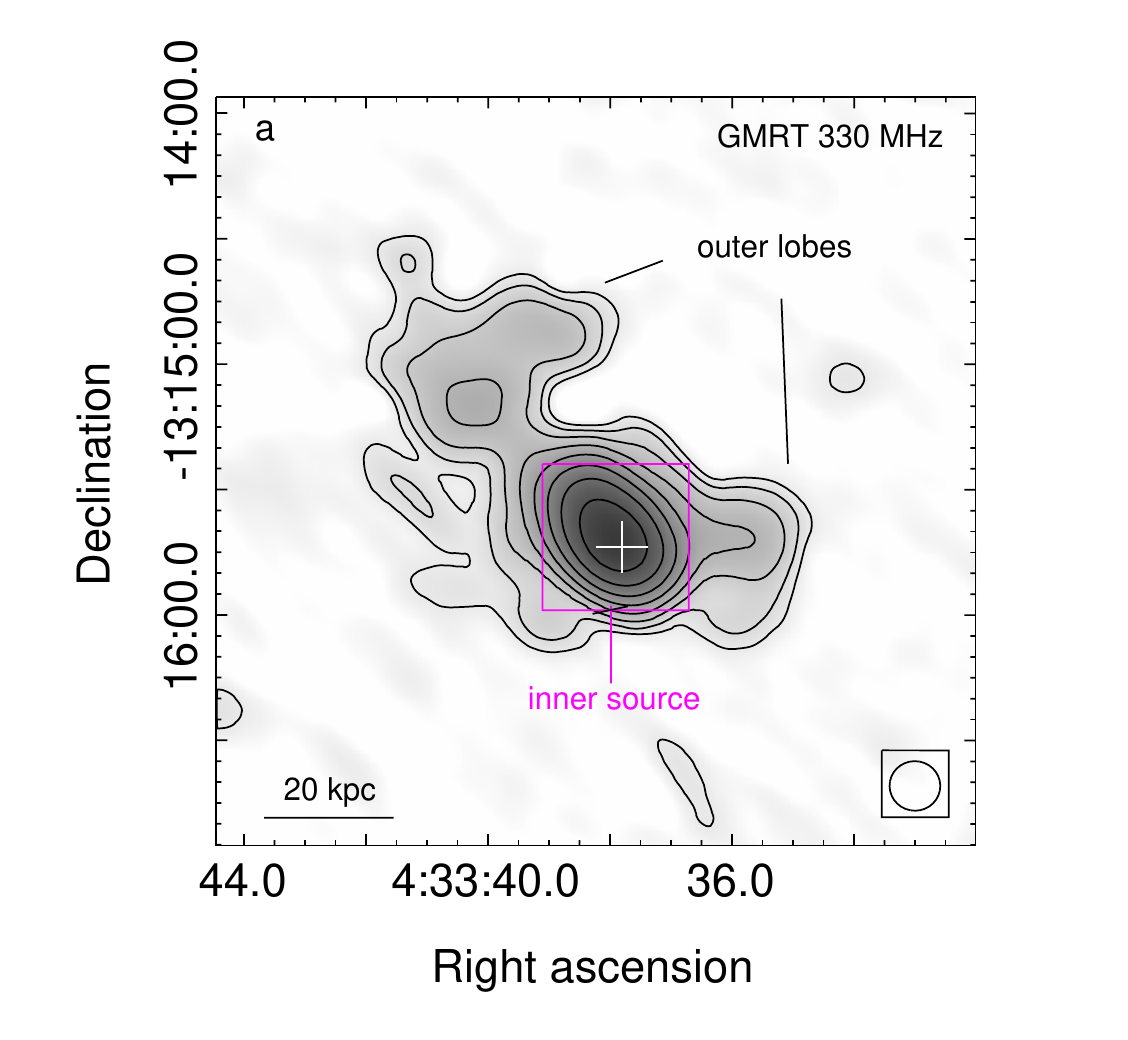}
\includegraphics[width=8.5cm]{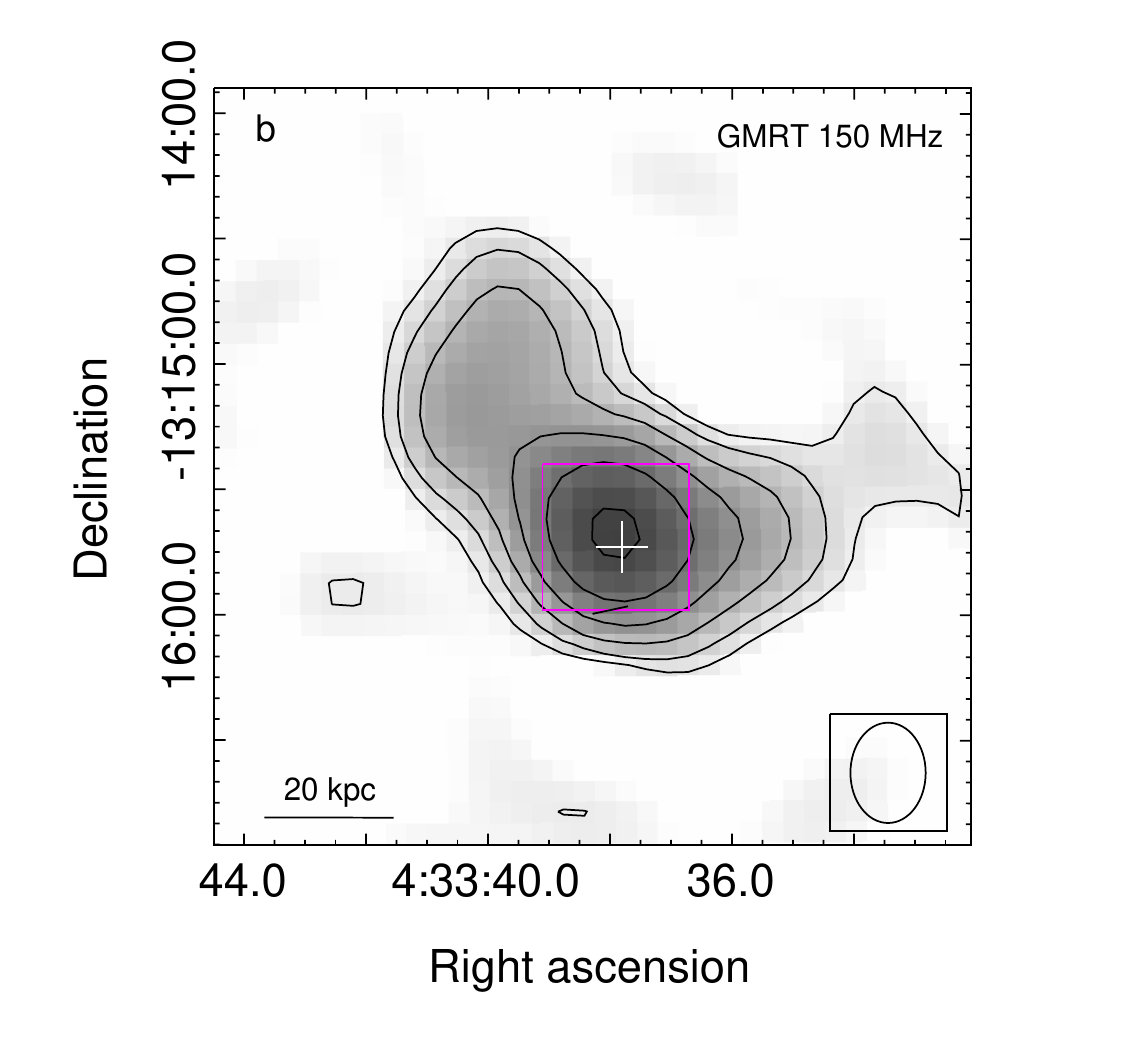}
\smallskip
\caption{a: GMRT image at 330 MHz (grayscale and contours). The image has been restored with a $12^{\prime\prime}$ circular beam and the noise is $1\sigma=0.5$ mJy/beam. 
b:  GMRT combined image at 150 MHz (grayscale and contours). 
The restoring beam is $24^{\prime\prime}\times18^{\prime\prime}$, in p.a. 0$^{\circ}$ and
$1\sigma=10$ mJy/beam. In both panels, contours are spaced by a factor of 2 
starting from $+3\sigma$. No contours at $-3\sigma$ are present in the portion of the image shown.
The boxed ellipse/circle shows the beam size. The magenta box marks the area occupied by the 
inner double source (see Fig.~\ref{fig:617}a). The white cross marks the optical peak (Fig.~\ref{fig:hst})}.
\label{fig:low}
\end{figure*}

\begin{table*}[ht!]
\renewcommand{\arraystretch}{1.2}
    \centering
    \setlength{\tabcolsep}{6pt}
    \caption{Total flux density of the radio galaxy in A496.}\label{tab:flux}
    \begin{tabular}{ccc|ccc|ccc}
         \hline
          {$\nu$} &
 {$S_{\nu,\,\rm tot}$} &
 {Reference} &
 {$\nu$} &
 {$S_{\nu,\,\rm tot}$} &
 {Reference} &
 {$\nu$} &
 {$S_{\nu,\,\rm tot}$} &
 {Reference} \\
 {(MHz)} & 
 {(mJy)} & 
 {code} &
 {(MHz)} & 
 {(mJy)} & 
 {code} &
 {(MHz)} & 
 {(mJy)} & 
 {code}  \\
         \hline
74      & $8292\pm1045$       & 1,2 & 174     & $2200\pm{177}$        & 3    &  3000    & $56.3\pm5.6$    & 1,5 \\
76      & $7992\pm{646}$        & 3   & 181     & $2064\pm{166}$        & 3    &  3000    & $48.5\pm4.9$    & 1,6 \\
84      & $7132\pm{575}$        & 3   & 189     & $1918\pm{155}$        & 3    &  3000    & $50.5\pm5.1$    & 1,7 \\
92      & $6255\pm{504}$        & 3   & 197     & $1882\pm{152}$        & 3    &   4710   & $63\pm3$        & 1 \\
99      & $5424\pm{437}$        & 3   & 204     & $1758\pm{142}$        & 3    &   4750   & $68\pm4$        & 8 \\
107     & $4996\pm{403}$        & 3   & 212     & $1572\pm{127}$        & 3    &   4848   & $54\pm3$        & 1 \\
115     & $4427\pm{357}$        & 3   & 220     & $1488\pm{120}$        & 3    &   4860   & $76\pm4$        & 1 \\
122     & $4034\pm{325}$        & 3   & 227     & $1422\pm{115}$        & 3    &   4860   & $67.5\pm3.3^a$  & 9 \\
130     & $3585\pm{289}$        & 3   & 330     & $822\pm82$         & 1    &   8640   & $51.0\pm2.5^a$  & 9 \\
143     & $2987\pm{241}$        & 3   & 340     & $574\pm86$         & 1    &  10700   & $46\pm8$        & 10 \\
150     & $3270\pm491$        & 1   & 617     & $273\pm22$         & 1    &  90000   & $11.4\pm3.6$    & 10 \\
151     & $2826\pm{228}$        & 3   & 888     & $207\pm14$         & 1, 4  &  150000  & $6.1\pm1.3$     & 10\\
158     & $2572\pm{207}$        & 3   & 1415    & $119\pm6$          & 1    &  150000  & $4.4\pm1.1$     & 10 \\
166     & $2314\pm{187}$        & 3   & 1490    & $109\pm5$          & 1    &          &                 &   \\                     
\hline
    \end{tabular}
\tablefoot{1: This work; 2: VLSSr; 3: GLEAM. {The errors represent the root sum squared of the statistical uncertainty reported in the GLEAM catalog and of the systematic uncertainty (8\%)}; 4: RACS-low; 5: VLASS 1.2; 6: VLASS 2.1; 7: VLASS 3.1; 8: \citet{andernach1998}; 9: \citet{2015MNRAS.453.1201H}; 10: \citet{2015MNRAS.453.1223H}. See Sect. \ref{sec:total_emission} for details. 
$^a$: No error is reported by the authors. A $5\%$ uncertainty is assumed here. }
\\
\end{table*}

\begin{table}
\renewcommand{\arraystretch}{1.2}
    \centering
    \caption{Flux densities of the compact component.}\label{tab:flux2}
    \setlength{\tabcolsep}{0.06\linewidth}
    \begin{tabular}{l|c|c}
         \hline
          {$\nu$} &
 {$S_{\nu,\,\rm compact}$} &
 {Reference} \\
 {(MHz)} & 
 {(mJy)} & 
 {code}\\
\hline
340                  & $70.4\pm10.6$   & 1 \\
617                  & $80.4\pm6.4$    & 1 \\  
1415                 & $69.4\pm3.5$    & 1  \\ 
1490                 & $77.2\pm2.0$    & 1  \\ 
3000                 & $56.3\pm5.6$    & 1,2 \\ 
3000                 & $48.5\pm4.9$    & 1,3 \\ 
3000                 & $50.5\pm5.1$    & 1,4 \\ 
4710                 & $60.6\pm3.0$    & 1  \\
4848                 & $54.2\pm2.7$    & 1 \\ 
4860                 & $76.0\pm3.8$    & 1 \\ 
4860                 & $66.6\pm3.3^{\rm a}$  & 5 \\  
8640                 & $48.8\pm2.3^{\rm a}$  & 5 \\  
10700                & $46.3\pm8.1$    & 6 \\ 
90000                & $11.4\pm3.6$    & 7 \\ 
150000               & $6.1\pm1.3$     & 7 \\ 
150000               & $4.4\pm1.1$     & 7 \\ 
\hline
    \end{tabular}
    \tablefoot{Column 1: This work; 2: VLASS 1.2; 3: VLASS 2.1; 4: VLASS 3.1;
    5: \citet{2015MNRAS.453.1201H}; 6: \citet{andernach1998}; 7: \citet{2015MNRAS.453.1223H}. See Sect. \ref{sec:compact} for details. $^{\rm a}$ No uncertainty is reported by the authors. An uncertainty of $5\%$ is assumed here.}
\end{table}

\begin{table}
\renewcommand{\arraystretch}{1.2}
    \centering
    \caption{Flux densities and spectral index of the lobes.}\label{tab:flux3}
    \setlength{\tabcolsep}{0.06\linewidth}
    \begin{tabular}{l|c|c}
        \hline
        {$\nu$} &
 {$S_{\nu,\,\rm lobes}$} &
 {$\alpha^{\rm a}$} \\
 {(MHz)} & 
 {(mJy)}  & \\
 \hline
 \multicolumn{3}{c}{{\large Inner lobes}} \\
\hline\noalign{\smallskip}
150  & $2440\pm366^{\rm b}$\\
330  & $732\pm73^{\rm b}$  \\  
340  & $504\pm76$ \\   
617  & $193\pm19$  \\    
1415 & $49.6\pm2.5$\\
1490 & $31.8\pm1.6$\\
4710 & $2.4\pm0.1$ & $2.0\pm0.1$ \\
\hline\noalign{\smallskip}
 \multicolumn{3}{c}{{\large Outer lobes}}\\
\hline\noalign{\smallskip}
150  & $830\pm125$\\
330  & $99\pm10$ & $2.7\pm0.2$  \\   
 \hline
    \end{tabular}
   \tablefoot{$^{\rm a}$ Spectral index measured between the lowest and highest frequency available.  $^{\rm b}$ Includes the central compact component. However, based on its spectrum (Fig.~\ref{fig:sp}), the contribution from the compact source is $\lax 4\%$ at 150 MHz and $\lax 10 \%$ at 330 MHz (see Sect. \ref{subsec:radiolobes} for details).}
\end{table}

\section{Complementary data}

\subsection{Chandra X-ray observations}
\label{sec:xray_obs}

A496 was observed by {\em Chandra} in 2000 for 18.9 ks (ObsID 931, ACIS-S, faint mode), in 2001 for 10 ks (ObsID 3361, ACIS-S, very faint mode), and in 2004 for 75 ks (ObsID 4976, ACIS-S, very faint mode). We reprocessed the event files in CIAO 4.14 \citep{2006SPIE.6270E..1VF}
using the Chandra Calibration Database (CALDB) 4.9.7 and standard procedures of calibrations (e.g., \citealt{2016ApJ...833...99W}), obtaining a cleaned exposure time of 65.3 ks (63\% of total time). To model the detector and sky background, we used the blank-sky data sets from the CALDB appropriate for the date of observation. The blank-sky data set was 
reprojected onto the sky using the aspect information and normalized using the 
ratio of the observed to blank-sky count rates in the 9.5--12 keV band. We obtained background-subtracted and exposure-corrected images of the cluster in the 0.5-4 keV and 0.5-7 keV energy band. Spectral analysis was performed in the 0.5 - 7 keV band using Xspec-v12.12.1, selecting the table of solar abundances of \citet{2009ARA&A..47..481A}. An absorption component (\texttt{tbabs}) was always included to account for Galactic absorption, with the column density fixed at $N_{\text{H}} = 4.3\times10^{20}$ cm$^{-2}$ \citep{2016A&A...594A.116H}, and the redshift fixed at $z=0.0329$.  

\subsection{VLT-MUSE observations}\label{subsec:musedata}
To complement our radio and X-ray analysis, we employ optical observations of A496 performed with the Very Large Telescope using the MUSE integral-field spectrograph (here we consider the observation 094.B-0592). The MUSE data were reduced using the MUSE pipeline 2.8.5 \citep{2014ASPC..485..451W}. The average seeing of the data is 0.8$^{\prime\prime}$. We fit the spectrum of each spaxel using the \texttt{PLATEFIT} code (\citealt{2004ApJ...613..898T,2004MNRAS.351.1151B}; see also e.g., \citealt{2019A&A...631A..22O} for a similar application) to derive a map of intensity and kinematics of the H$\alpha$ emission line, which is sensitive to the extended warm gas nebula surrounding the central dominant (cD) galaxy.

\section{Radio images of A496}\label{sec:img}

In this section, we present our radio images of the central source in A496 
at multiple angular resolution and frequencies, which allow us to identify 
different components of emission on different spatial scales.

\subsection{Compact component and inner lobes}\label{sec:inner}

Fig. \ref{fig:hst} presents our high resolution ($\sim 2^{\prime\prime}$) image at 1.5 GHz (blue colour and contours) overlaid on a Hubble Space Telescope (HST) 
Wide Field and Planetary Camera 2 (WFPC2) image of the cluster central 
galaxy (Observation ID U62G1902R). The radio emission (hereafter referred 
to as inner source) consists of a bright central and compact component, 
coincident with the galaxy optical peak, and a pair of asymmetric, 
diffuse and very faint radio lobes. The lobes extend $\sim10$ kpc toward 
northeast (NE) and $\sim5$ kpc in the southwest (SW) direction. No defined, kpc-scale 
jets are visible within their diffuse emission. 

In Fig. \ref{fig:617} we show the VLITE 340 MHz, GMRT 617 MHz, VLA-B 1.4 GHz 
and VLA-C 4.7 GHz images of the inner source. The central component and inner 
lobes are detected on a similar scale of $\sim 25$ kpc in panels a, b and c. 
At 4.7 GHz (d), the emission is clearly dominated by the compact component 
and only part of the NE inner lobe is detected.

\subsection{Outer radio lobes}

Fig. \ref{fig:low} presents our low-frequency {\em GMRT} images 
at 150 MHz\footnote{An image at 150 MHz from the same data sets analyzed here is available from the TIFR GMRT Sky Survey Alternative Data Release \citep[TGSS-ADR,][]{2017A&A...598A..78I}, where the radio source is detected with a similar morphology and flux density as measured in our image.} and 330 MHz. The images are shown on the same spatial scale. 
As a reference, the magenta box marks the area occupied by the inner lobes and compact component (Fig.~\ref{fig:617}a). 
A pair of large-scale radio lobes are detected in both images. The lobe axis is
close to that of the inner lobes in Fig. \ref{fig:617}, however both lobes seem 
to slightly bend at large distance from the center. The total extent of the source 
at these low frequencies is $\sim 100$ kpc.

%
\begin{figure*}[ht!]
\centering
\includegraphics[width=\linewidth]{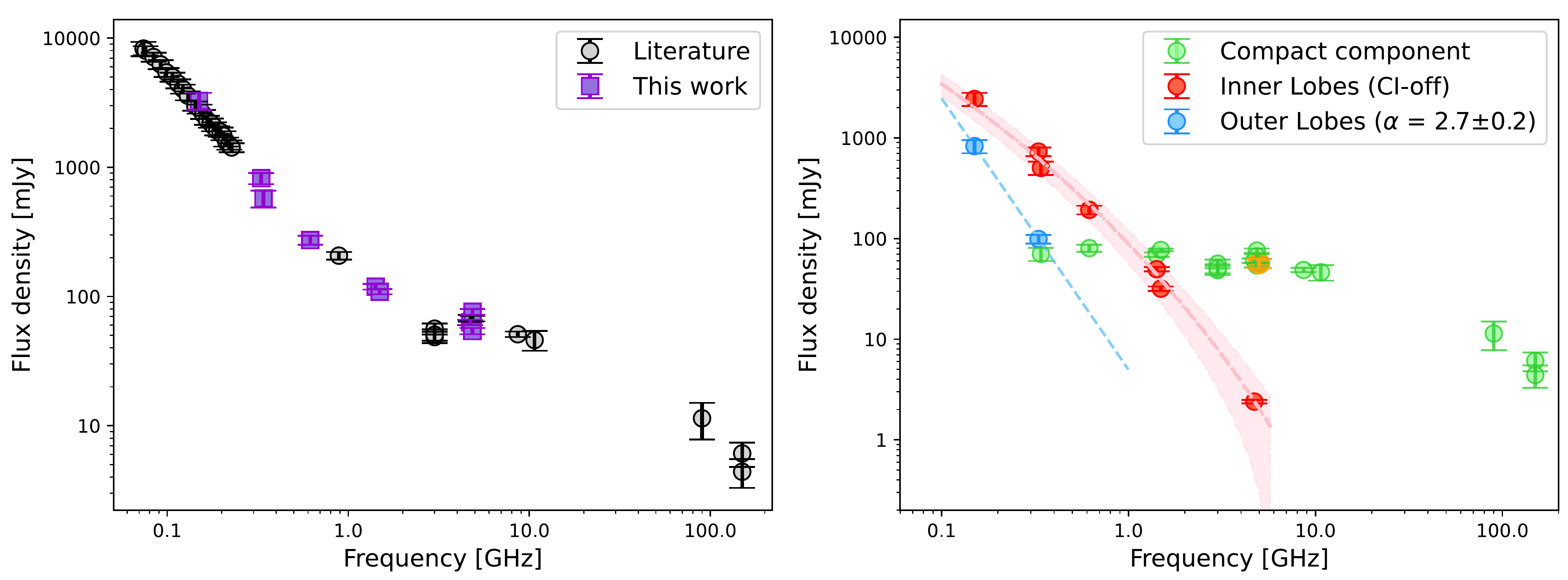}
\smallskip
\caption{{\em Left:} Integrated radio spectrum of the total radio emission at the center of A496. Our flux density measurements are shown as purple filled squares. {Catalog and literature values from Table~\ref{tab:flux} are shown in black}. {\em Right:} Radio spectra of the individual components. The blue dashed line shows the spectral slope of the outer lobes. The dashed pink line and shaded area represent the best fit and associated uncertainty with a CI-off model to the spectrum of the inner lobes. The VLBA 5 GHz flux density from \citet{2024ApJ...961..134U} is plotted with a gold circle.}
 \label{fig:sp}
\end{figure*}

\begin{figure}
    \centering
    \includegraphics[width=\linewidth]{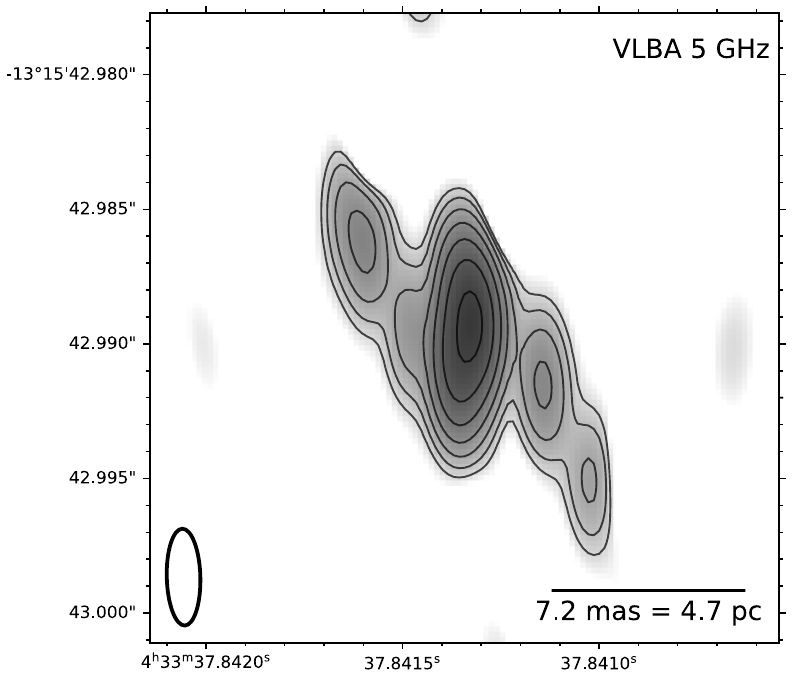}
    \caption{{VLBA image at 5 GHz of the compact component in A496, at a resolution of $3.5\times1.2$~mas and with noise $\sigma = 60$~$\mu$Jy/beam. Contours start at $5\sigma$ and increase by a factor of 2.}}
    \label{fig:VLBA}
\end{figure}

\section{Radio flux densities and spectral analysis}\label{sec:flux}

In this section we measure the total flux densities and radio spectrum 
of the central source in A496 and of its different components 
using the images listed in Table~\ref{tab:images}. 
We complement our measurements with flux densities from the literature and from radio surveys, including the VLA Low-frequency Sky Survey Redux \citep[VLSSr,][]{2014MNRAS.440..327L} at 74 MHz, the Galactic and Extragalactic All-Sky MWA survey \citep[GLEAM,][]{2017MNRAS.464.1146H} at 76--227 MHz, the Rapid ASKAP Continuum Survey \cite[RACS,][]{2020PASA...37...48M, 2021PASA...38...58H} at 888 MHz, and the VLA Sky Survey \citep[VLASS,][]{2020PASP..132c5001L} at 3 GHz.
For a proper comparison, we have re--scaled all the flux density values used in this paper to a common flux density scale. We adopted the \cite{2017ApJS..230....7P} scale, which is valid between 50 MHz and 50 GHz, and used appropriate scaling factors based on \cite{2017ApJS..230....7P}. 
Differences between the original and \cite{2017ApJS..230....7P} scales are 
estimated to be within $5\%$.

\subsection{Total emission}\label{sec:total_emission}

We measured the total radio emission at the cluster center by integrating 
our images within a circular region of $1^{\prime}.5$ radius ($\sim$ 60 kpc), centered on the 
radio peak (RA$_{\rm J200}=$04h33m38s, DEC$_{\rm J2002}=-13^{\circ}15^{\prime}43^{\prime\prime}$). 
For the images at 4.8 GHz and 4.9 GHz in B and A configurations, where only 
the central compact component is detected, the total flux density was measured 
via Gaussian fit on the images. Our measurements are summarized in Table~\ref{tab:flux}. 
Errors were computed including the image rms and flux calibration uncertainty. 
The table includes complementary flux densities from the 74 MHz VLSSr and 888 
MHz RACS images (measured within the same $r=1^{\prime}.5$ region), catalogue values from the GLEAM survey, and higher-frequency measurements reported in the literature with the Australia Telescope Compact Array \citep{2015MNRAS.453.1201H}, Effelsberg \citep{andernach1998}, CARMA and GISMO \citep{2015MNRAS.453.1223H}. 
The 3 GHz measurements are from the VLASS quick-look images from Epoch 1.2, 2.1 and 3.1, where only 
the compact source is detected. The corresponding flux densities were calculated using a Gaussian 
fit to the source on the quick-look images and corrected for a low-flux density bias 
of $15\%$ (VLASS 1.2) and $8\%$ (VLASS 2.1 and 3.1) that affects faint sources, as reported 
in the NRAO Quick-Look Image web page\footnote{\url{https://science.nrao.edu/vlass/data-access/vlass-epoch-1-quick-look-users-guide}}. A total uncertainty of $10\%$ was assumed for the VLASS values. 
As mentioned above, all flux density values are on the same \cite{2017ApJS..230....7P} scale, 
with the exception of the measurements at 90 GHz (CARMA) and 150 GHz (GISMO).

The spectrum of the whole radio emission is shown in Fig. \ref{fig:sp} ({\em left}) 
with data points from our images highlighted in violet. A similar spectrum, but with 
fewer data points, was presented by \cite{2015MNRAS.453.1223H}. 
The overall shape of the spectrum is complex and clearly inconsistent with a simple 
power-law behavior. The spectral index is very steep ($\alpha\sim1.5$) at frequencies 
lower than 1.4 GHz, it flattens to $\alpha\sim 0.4$ between 1.4 GHz and 10.7 GHz, 
and then it steepens again to $\alpha\sim 0.9$ between 10.7 GHz and 150 GHz. 
The flattening and curving of the spectrum above 1.4 GHz suggests the presence of an 
active component that dominates the emission at high frequencies. The steepening 
below 1.4 GHz suggests, instead, the presence of an aged, extended component, that 
emerges predominantly at low frequencies. To investigate further these spectral 
features, we examined the spectrum of the individual components in the system.

\subsection{Central compact component}\label{sec:compact}

We measured the flux density of the central compact component (Figs.~\ref{fig:hst} and \ref{fig:617}) 
by fitting the source with a Gaussian model. At 340 MHz and 617 MHz, we used images made 
using only baselines longer than 20 k$\lambda$ (sensitive to structures smaller than a few kpc). 
Our flux densities are summarized in Table~\ref{tab:flux2}, 
which also includes the 3 GHz flux densities from the VLASS 1.2, 2.1 and 3.1 quick-look images 
(see Sect.~\ref{sec:total_emission}) and higher-frequency measurements from the literature.  
The spectrum of the central component in the 340 MHz--150 GHz interval is shown 
in Fig.~\ref{fig:sp} ({\em right}) as green data points. The central source has a steep, 
high-frequency spectrum ($\alpha \sim 0.7$ above 5 GHz) and a flat, possibly inverted, 
spectrum below 5 GHz. This is similar to the spectral behavior of GHz-Peaked Spectrum (GPS) 
radio sources 
(e.g., \citealt{1997A&A...325..943S,2021A&ARv..29....3O}),
which are interpreted as an early stage in the evolution of a radio galaxy. The similarity suggests that 
the central compact source represents the onset of a new phase of activity of the central AGN.
\\This is consistent with the detection of a radio core in Very Long Baseline Array (VLBA) archival 
data at 5~GHz (project BE056), as recently shown by \citet{2024ApJ...961..134U}. In the VLBA image that we show in {Fig. \ref{fig:VLBA}}, 
twin jets are seen emanating from the core in the NE-SW direction (position angle of about $140^{\circ}$), 
and the source has a largest linear size of 15~mas (10 parsecs). The core has a flux density at 5~GHz of $44.3\pm4.3$~mJy, whereas the jets have a combined flux density of $12.6\pm1.3$~mJy. The total of about $57\pm6$~mJy represents $\sim$100\% of the 
flux density detected at kpc scales from the VLA at around 5~GHz (Fig.~\ref{fig:617}d, Fig.~\ref{fig:sp} {\em right}, and Table~\ref{tab:flux2}). 
This supports the scenario of the central compact source being a recently renewed radio galaxy.

%
\begin{figure*}[ht!]
\centering
\includegraphics[width=8cm]{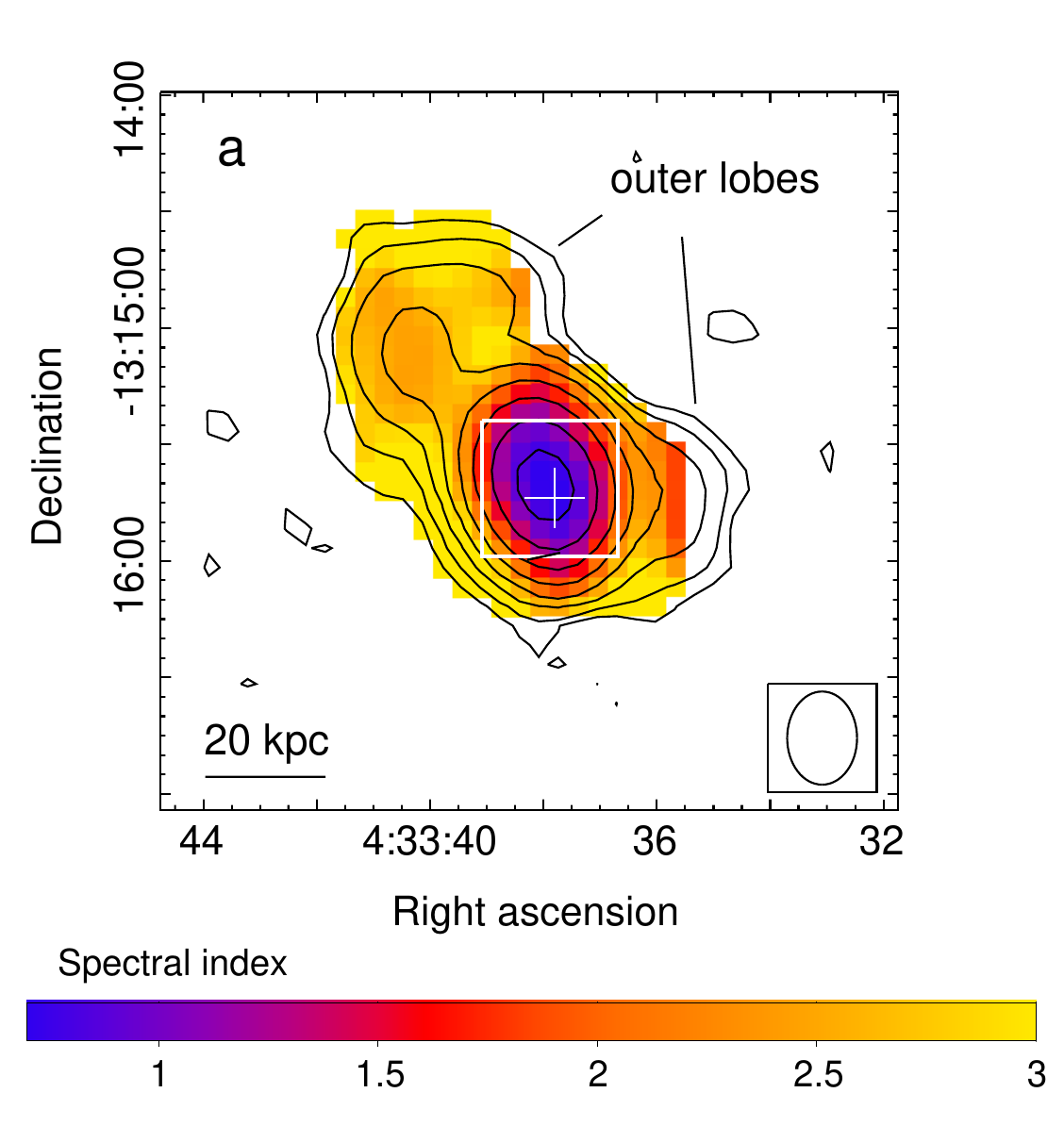}
\includegraphics[width=8cm]{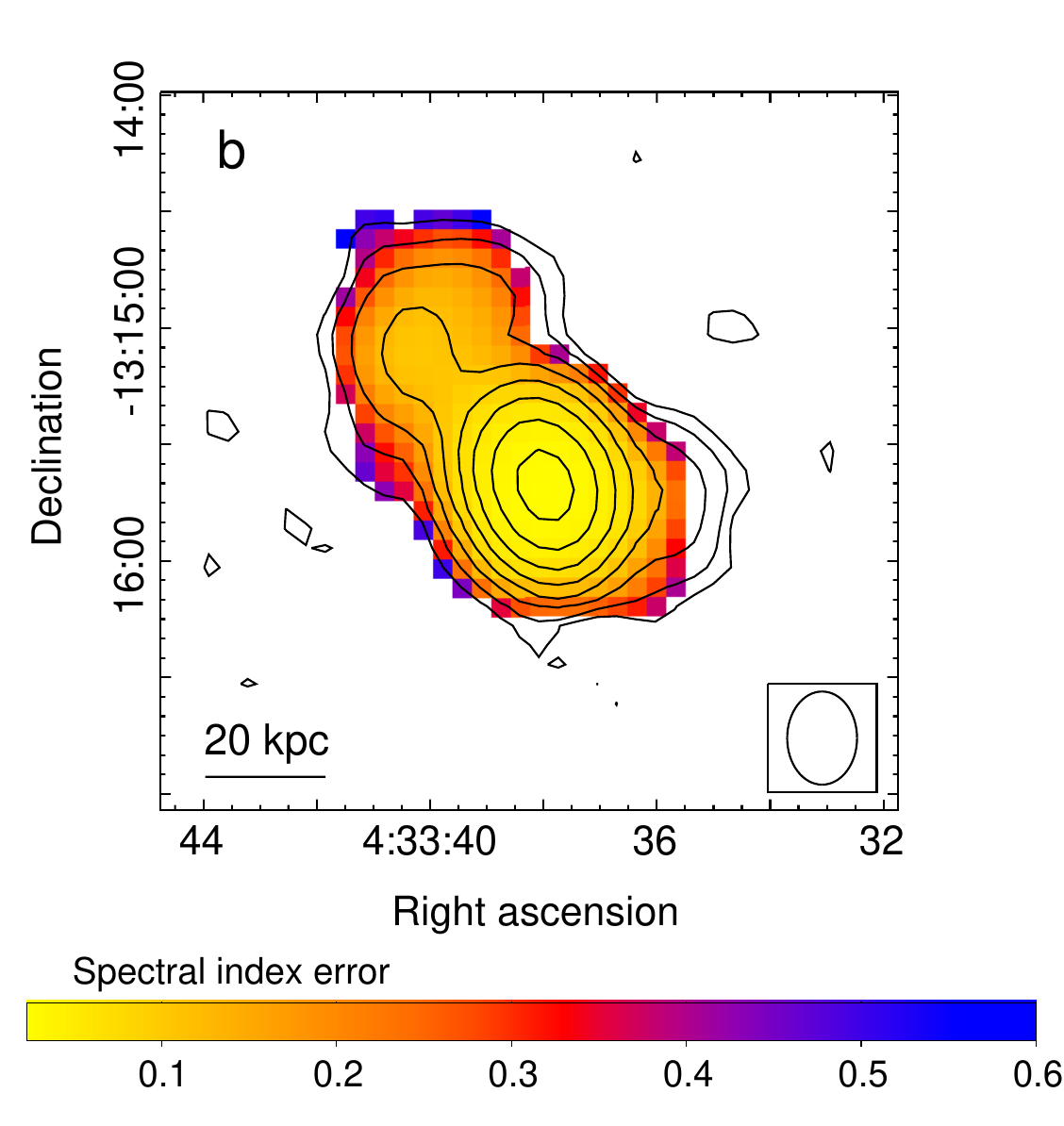}
\smallskip
\caption{Color-scale image of the spectral index distribution between 150 MHz and 330 MHz
(a) and associated error map (b), computed from primary beam corrected images with same uv range and beam of $24^{\prime\prime}\times18^{\prime\prime}$. The spectral index was calculated in each pixel where the surface brightness is above the $3\sigma$ level in both images. Overlaid are the 330 MHz
contours, spaced by a factor of 2 from $3\sigma=3$ mJy/beam. The white box marks the region occupied by the inner double source (Fig.~\ref{fig:617}a). The white cross marks the optical peak (Fig.~\ref{fig:hst}).}
\label{fig:spix}
\end{figure*}

\subsection{Radio lobes}\label{subsec:radiolobes}

The flux densities of the inner lobes are summarized in Table~\ref{tab:flux3}, along with the associated uncertainties and with the observed spectral index measured between the lowest and highest frequencies available. At frequencies $\ge 340$ MHz, the lobe flux density was estimated by subtracting the contribution of the compact component (Table~\ref{tab:flux2}) from the total emission in the area shown in Fig.~\ref{fig:617}a. At 150 MHz and 330 MHz, the angular resolution of our images does not allow us to separate the compact source from the surrounding extended emission. Therefore, in Table~\ref{tab:flux3}, we report the total flux density (inner lobes + compact source) measured within the magenta box in Fig. \ref{fig:low}{\em a}. Based on its flat spectrum at GHz frequencies (Fig.~\ref{fig:sp}, {\em right}), we estimate that the contribution of the compact component is at most $\sim 10\%$ at 330 MHz, and less than $\sim 4\%$ at 150 MHz. 

%
\begin{figure*}[ht!]
\centering
\includegraphics[width=7.5cm]{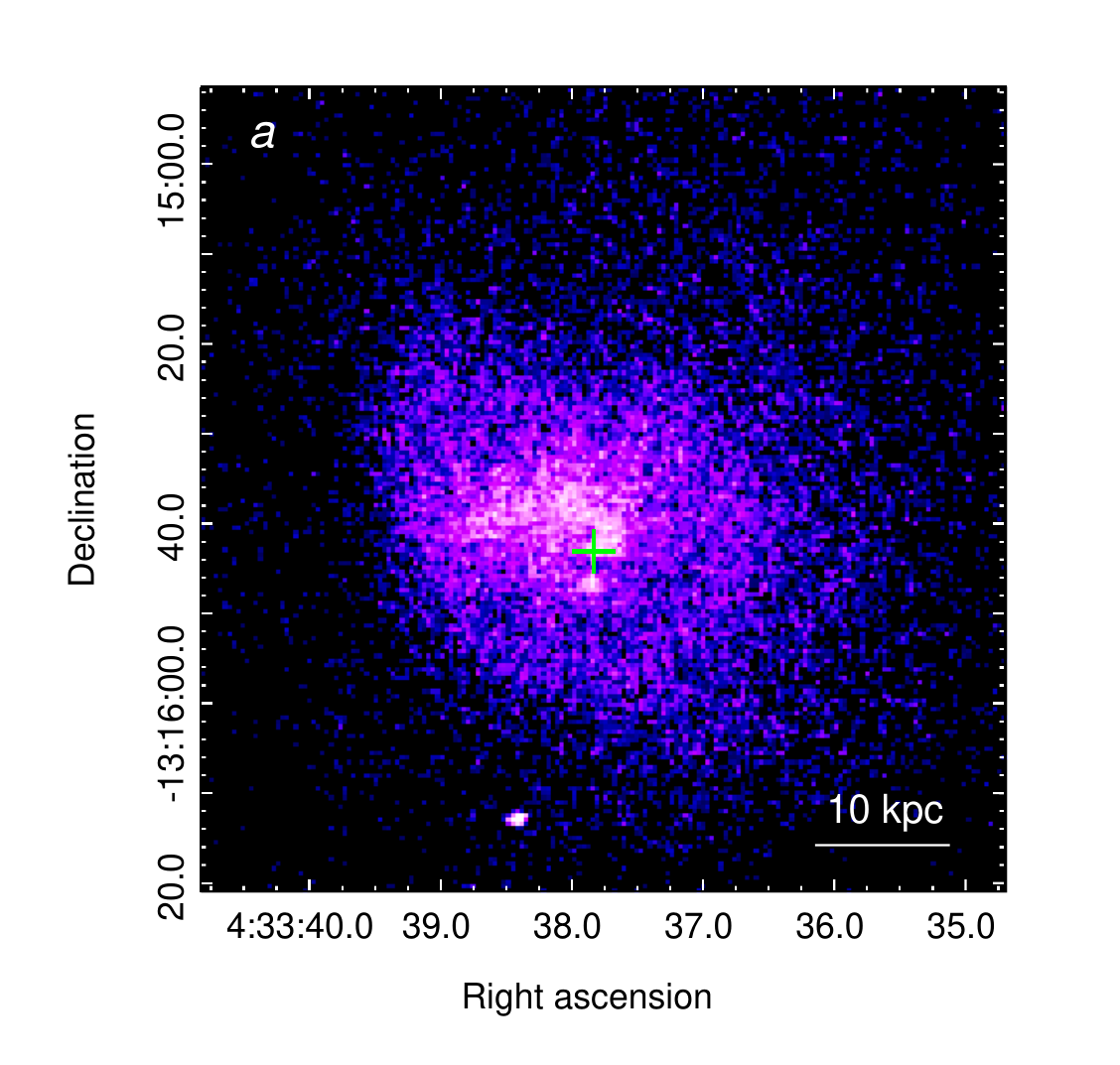} 
\hspace{0.25cm}
\includegraphics[width=7cm]{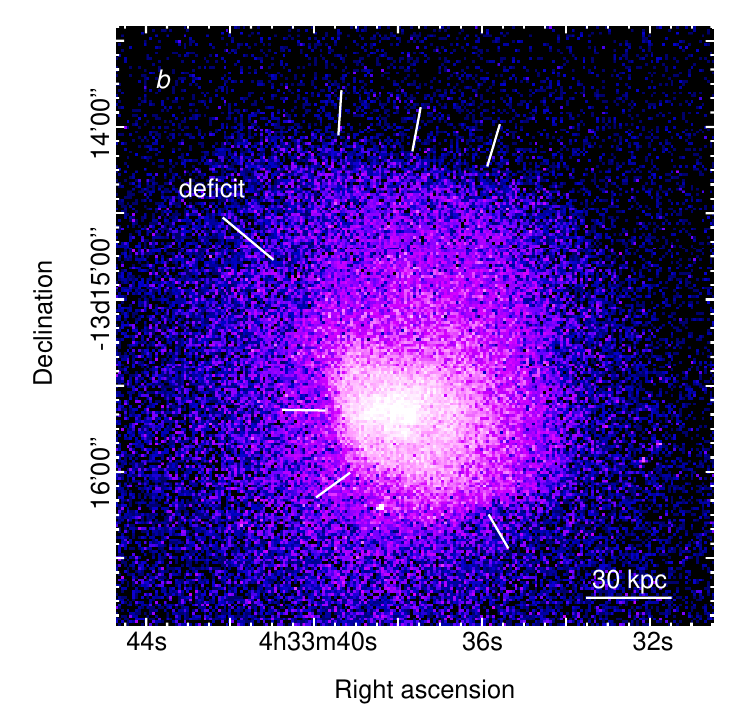}
\includegraphics[width=7cm]{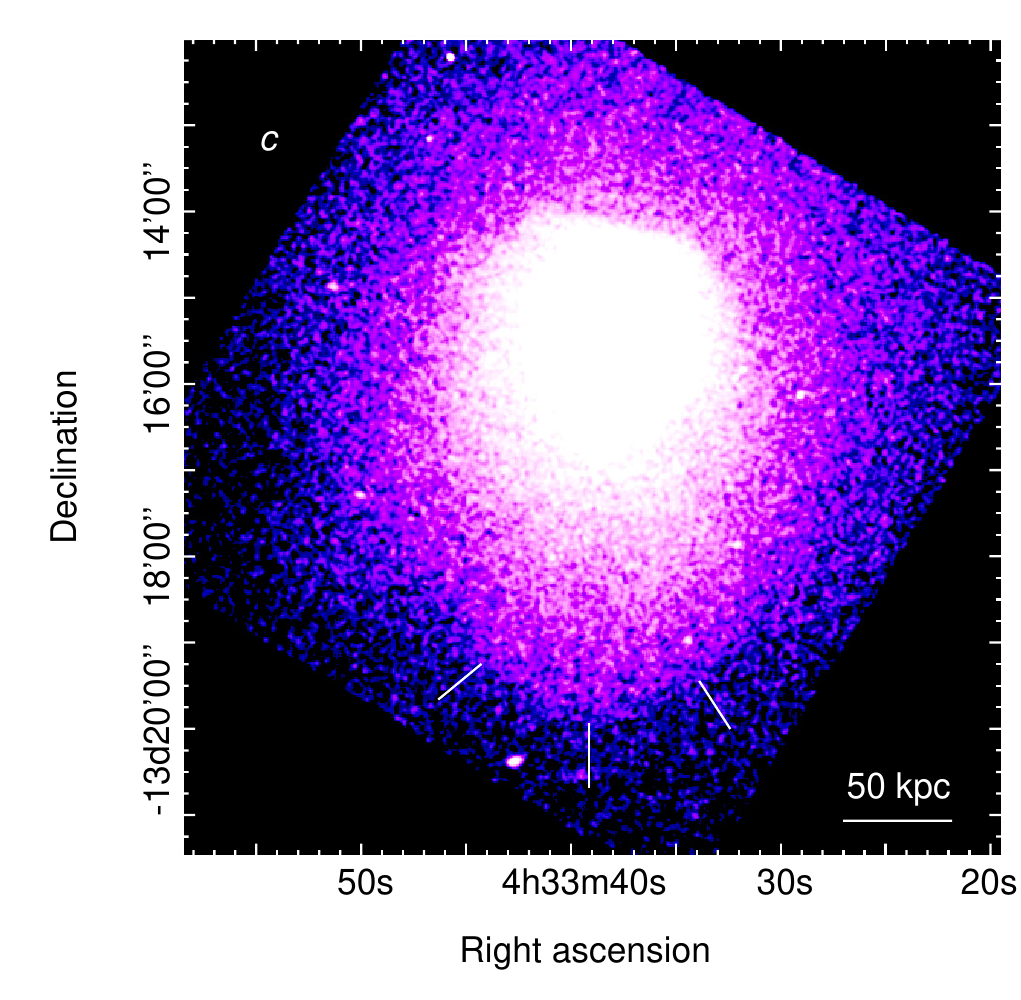}
\hspace{0.2cm}
\includegraphics[width=7cm]{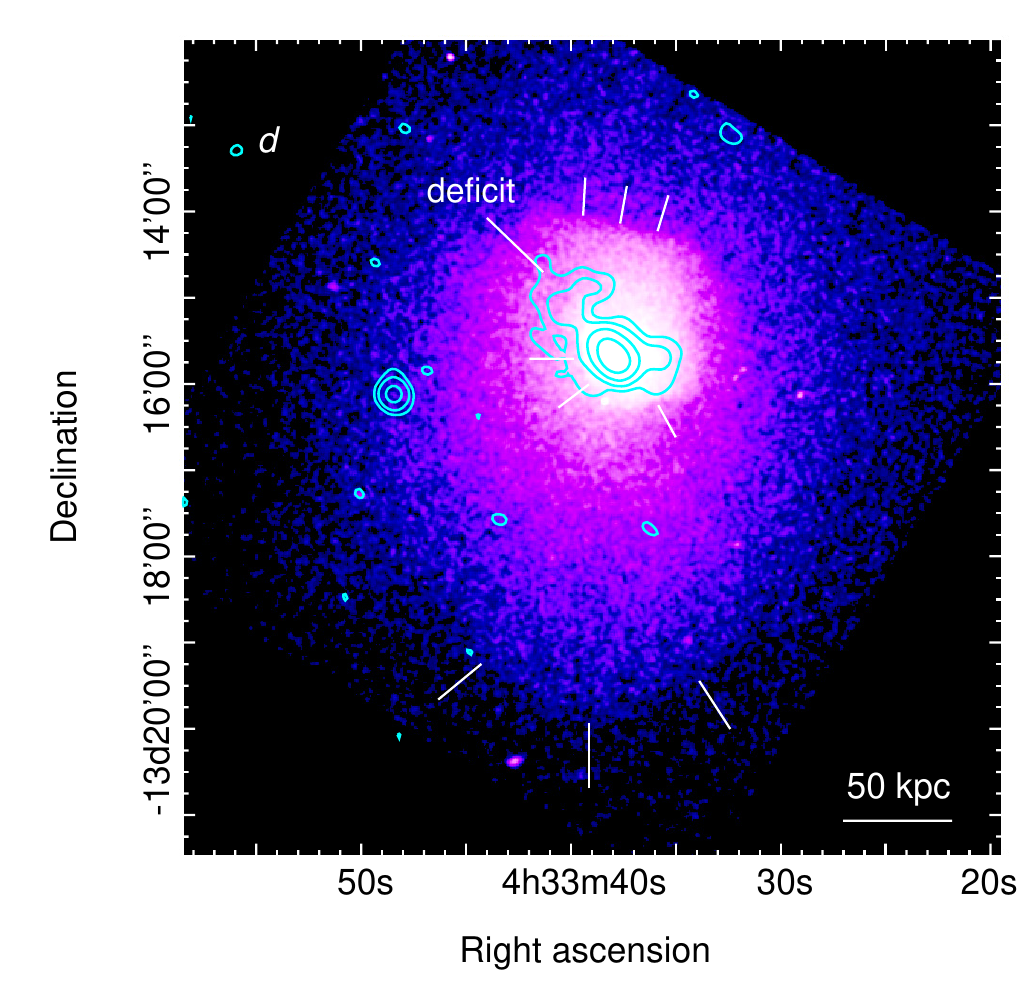}
\smallskip
\caption{({\em a}) {\em Chandra} 0.5-4 keV image of the central 60 kpc $\times$ 60 kpc region of A496, smoothed
with a $\sigma=0^{\prime\prime}.5$ Gaussian. The green cross marks the location of the cD 
galaxy. ({\em b}) {\em Chandra} 0.5-4 keV image of the central 140 kpc $\times$ 140 kpc region. 
White lines mark the position of the inner cold fronts. The image hints to the presence
of a large decifit NE of the center. ({\em c}) {\em Chandra} 0.5-4 keV image of the whole cool core,
smoothed with a $\sigma=1^{\prime\prime}$ Gaussian. White lines mark the position of the outermost cold front.
({\em d}) {\em GMRT} 330 MHz contours from Fig.~\ref{fig:low}({\em a}) at levels of 
1.8, 7.2, 28.8 and 115.2 mJy/beam ($12^{\prime\prime}$ resolution), overlaid on the {\em Chandra} 0.5-4 keV image, 
smoothed with a $\sigma=1^{\prime\prime}$ Gaussian.}
\label{fig:chandra}
\end{figure*}
\begin{figure*}[ht!]
    \centering
    \includegraphics[width=\linewidth]{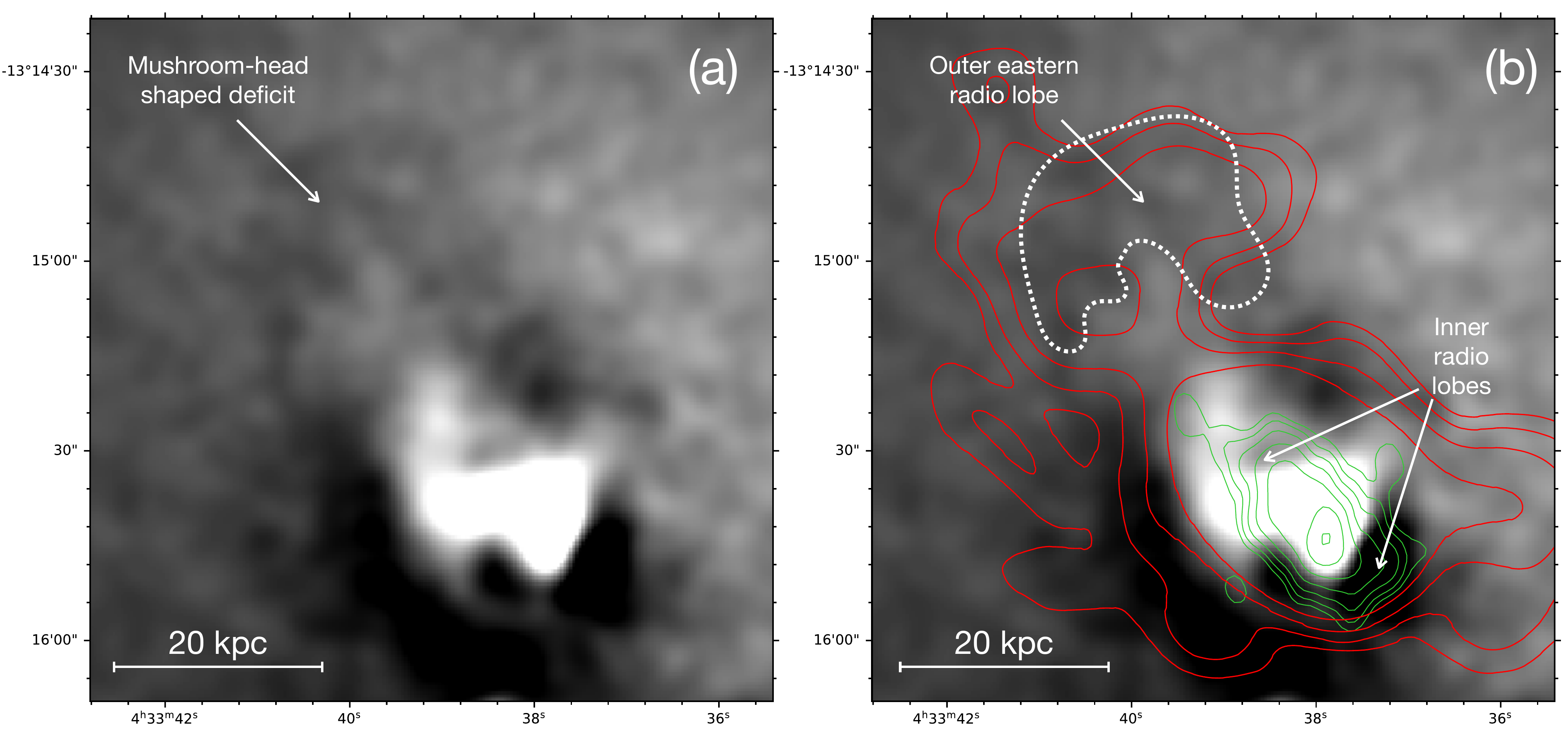}
    \caption{(a) {\it Chandra} unsharp masked image of A496, obtained by subtracting a 10$^{\prime\prime}$-$\sigma$ smoothed image from a 2$^{\prime\prime}$-$\sigma$ smoothed one. The image has been further smoothed with a 4$^{\prime\prime}$-$\sigma$ Gaussian. (b) Same as panel (a), with red contours from the GMRT at 330 MHz at levels of 1.5, 3, 6, 12 mJy/beam. Green contours are from the VLITE at 340 MHz at levels of 2, 4, 8, 16, 32, 64, 128 mJy/beam. The dashed white region shows the mushroom head-shaped cavity.}
    \label{fig:chandra-resid}
\end{figure*}

The outer lobes are clearly detected only in our images at 150 MHz and 330 MHz. We measured their flux density at these frequencies
as difference between total emission (Table~\ref{tab:flux}) and inner emission (inner lobes and compact source; Table~\ref{tab:flux3}). 
We obtained $830\pm115$ mJy at 150 MHz and $99\pm10$ mJy
at 330 MHz (Table~\ref{tab:flux3}).

In Fig.~\ref{fig:sp} ({\em right}) , the spectra of the inner and outer lobes are shown in red and blue, respectively. The spectrum of the inner lobes has a steep slope of $\alpha=1.5\pm0.1$ between 150 MHz and 617 MHz and it further steepens to $\alpha=2.3\pm0.1$ at higher frequencies. 
The outer lobes have an even steeper spectrum with $\alpha=2.7\pm0.2$ in the 150--330 MHz interval. 
These steep indices suggest that both pairs of lobes are old and possibly associated with former cycle(s) of activity of the central galaxy.

Moreover, the diffuse morphology of the inner lobes and the presence of a renewed outburst on pc scales indicate that the inner lobes are no-longer powered by active jets. Thus, we fitted their spectrum using SYNCHROFIT\footnote{\url{https://github.com/synchrofit/synchrofit}.} \citep{2022MNRAS.514.3466Q} with a CI-off model, which assumes an initial phase of electron injection at a constant rate by the nuclear source (continuous injection, 
CI), followed by a dying phase during which the radio emission fades (e.g., \citealt{2011A&A...526A.148M}). It is assumed a uniform magnetic field within the source and and an isotropic distribution of the pitch angle of the radiating electrons. Prior to fitting, we set the magnetic field to the equipartition value of $B_{eq} = 22$~$\mu$G estimated from the 1.4 GHz map\footnote{We assumed a filling factor of 1 and zero protons ($k=0$).} of Fig. \ref{fig:617} and using the equations reported in \citet{2004IJMPD..13.1549G}. The resulting best fit for the spectrum of the inner lobes (pink line in Fig. \ref{fig:sp}) gives a total age of $t_{tot} = 26 \pm 4$~Myr, with an active phase of duration $t_{on} = 20 \pm 3$~Myr and a passive phase of duration $t_{off} = 6\pm 1$~Myr. 
\\Given the availability of two points only in the spectrum of the outer lobes, we are unable to derive their synchrotron ages. In any case, their ultra-steep spectrum between 150 and 330 MHz (with $\alpha = 2.7\pm0.2$) strongly suggests that they are older than the inner lobes. We can derive an approximate radiative age using the following equation (e.g., \citealt{2014NJPh...16d5001E}): 
\begin{equation}
    t_{\text{rad}}[\text{Myr}] = \frac{1590\sqrt{B}}{(B^{2} + B_{CMB}^{2})\sqrt{1+z}}\sqrt{\frac{(\alpha-\Gamma)\text{ln}(\frac{\nu_{2}}{\nu_{1}})}{\nu_{2} - \nu_{1}}}
    \label{eq:trad}
\end{equation}
\noindent where $B$ and $B_{CMB}=3.25(1+z)^{2}$ are the source and the Cosmic Microwave Background magnetic fields, respectively ($\mu$G), and $\Gamma$ is the injection index (assumed equal to 0.7, e.g., \citealt{2014NJPh...16d5001E}). For $B\sim1 - 20\,\,\,\mu$G we find $t_{\text{rad}}\sim50-300$~Myr. We caution that this estimate is an approximation, and we can at most argue that the radiative age of the outer lobes is of the order of $\sim10^{8}$~yr.


\subsection{Spectral index image}

A spectral index image in the low-frequency interval 150--330 MHz 
is shown in Fig.~\ref{fig:spix},
along with the associated error map. The image was obtained by comparing a pair of 
primary-beam corrected images produced with the same cell size, $uv$ range 
($0.06-12.2$ k$\lambda$), and restoring beam of $24^{\prime\prime}\times 18^{\prime\prime}$. Only pixels with signal-to-noise ratio 
of 3 or larger were considered.

The image confirms that the spectral index in the outer lobes is very steep ($\alpha \sim 2.5-3$, with a typical spectral index uncertainty of $\sim$0.2).
The central region, that encloses the inner lobes (white box), is characterized by a peak with $\alpha\sim 0.7$, surrounded by steeper-spectrum emission with values up to $\sim 2$.

\section{Connection between radio, X--ray, and H$\alpha$ emission}\label{sec:connectionratioX}

A496 is known for its striking four concentric cold fronts, three of them located 
within the central $r\sim70$ kpc region \citep{2003ApJ...583L..13D,2007ApJ...671..181D} 
and the most distant one at $\sim 160$ kpc from the center (\citealt{2014A&A...570A.117G}; see also \citealt{2006PASJ...58..703T}). Here, we focus on the details of the 
X-ray emission within the cooling region ($r\sim70$~kpc) and its relation to the extended 
radio features associated with the cD galaxy.

In Fig.~\ref{fig:chandra} we show the X-ray emission detected by {\em Chandra} at increasing distance from the center. The core structure inside the 
innermost cold front, shown in panel ({\em a}), is characterized by a bright curved 
feature -- possibly the tip of the sloshing spiral seen on larger scale -- and a number of 
small surface brightness depressions, distributed around the position of the cD 
galaxy (marked by the green cross).  
On larger scale ({\em b}), the most prominent 
feature is the spiral structure traced by the cold fronts (marked by white lines). 
A region of X-ray emission deficit is visible to the NE of the center. Zooming out ({\em c}), an additional sharp surface brightness discontinuity is visible near the edge of the whole image. This corresponds to the outermost cold front found with {\em XMM-Newton} by \citet{2014A&A...570A.117G}. The location of the 330 MHz radio emission with respect 
to the X-ray surface brightness distribution of the cool core is shown in panel {\em d}. 
\begin{figure*}[ht!]
    \centering
    \includegraphics[width=0.95\linewidth]{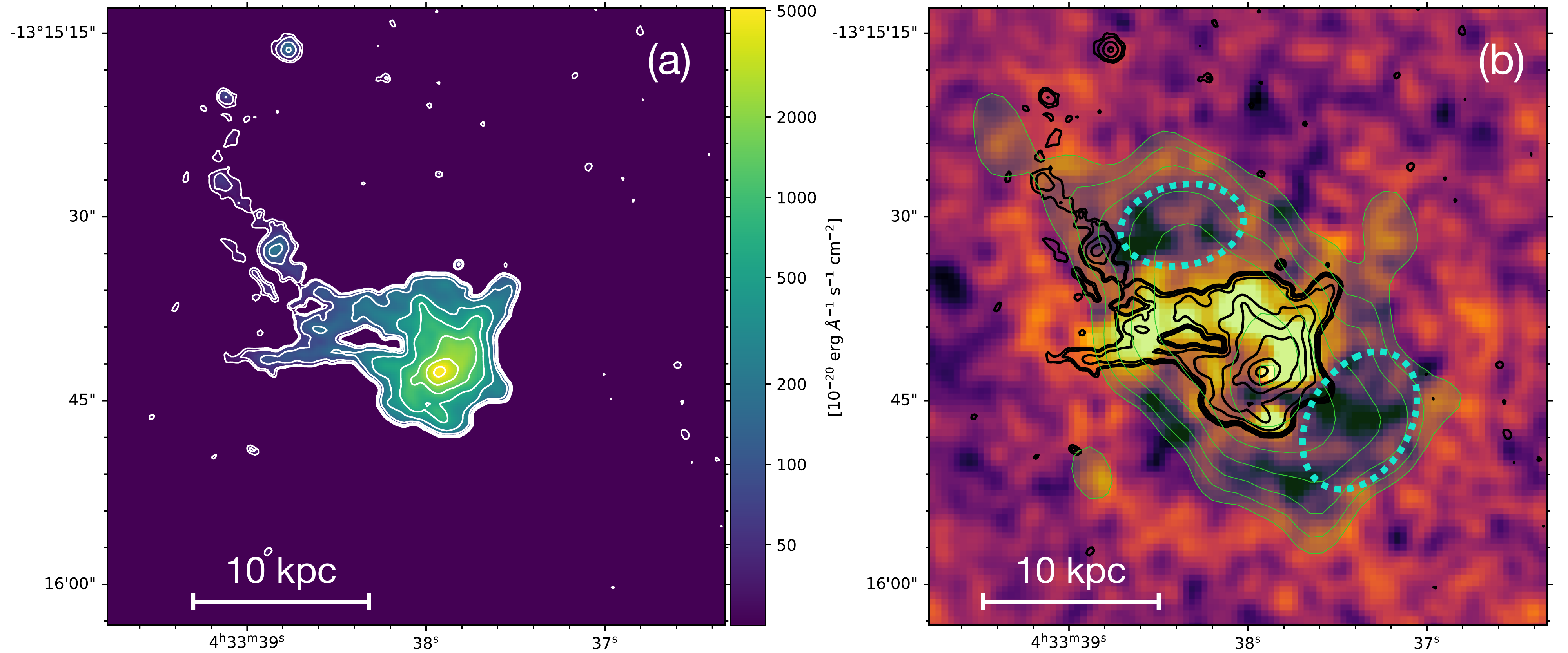}
    \caption{(a) MUSE image of the H$\alpha$ line intensity, with overlaid white contours. (b) {\it Chandra} unsharp masked image of A496, obtained by subtracting a 10$^{\prime\prime}$-$\sigma$ smoothed image from a 2$^{\prime\prime}$-$\sigma$ smoothed one. Green contours are from the VLITE at 340 MHz at levels of 2, 4, 8, 16, 32, 64, 128 mJy/beam, while black contours are from panel (a). Dashed cyan ellipses show the position of the cavities excavated by the inner lobes (see Sect. \ref{sec:connectionratioX})}.
    \label{fig:chandra-condensation}
\end{figure*}
The radio source occupies only a small fraction of the cool core and
is mostly coincident with the brightest X-ray emission north of the two innermost cold 
fronts. 

To examine the innermost 70 kpc structures in more detail, we produced an unsharp masked image by smoothing the 0.5 - 7 keV {\it Chandra} images by two Gaussians of $\sigma=2^{\prime\prime}$ and $\sigma=10^{\prime\prime}$. The subtraction of the lightly-smoothed image from the heavily-smoothed one produced the unsharp masked image shown in Fig. \ref{fig:chandra-resid}. {The unsharp masked image highlights the large deficit identified in Fig.~\ref{fig:chandra}{\em b}, that is filled by the NE outer lobe. This suggests that the X-ray deficit is an X-ray cavity. Upon inspection, the deficit displays a mushroom head-shaped morphology, which is also strikingly consistent with the morphology of the radio lobe detected at 330~MHz (red contours in panel b; see also Fig. \ref{fig:low}a).} Theoretical arguments predict X-ray cavities to assume a torus or mushroom-head shape while buoyantly rising through the cluster potential. As the bubble rises, a trunk of dense gas is entrained in its wake, which penetrates it from below and gives it a mushroom shape (e.g., \citealt{2007A&A...464..143G}). Interestingly, this is consistent with the morphology we observe, since there is a region of enhanced X-ray emission right "behind" the cavity that may represent the entrained wake (see Fig. \ref{fig:chandra}b, \ref{fig:chandra-resid}, and \ref{fig:chandra-new}a). As visible in Fig. \ref{fig:chandra-resid}{\em b}, two additional X-ray depressions are clearly coincident with the inner radio lobes 
detected at 340 MHz. These features are thus interpreted as real cavities created by displacement of the thermal gas by the inner radio lobes. By comparing the X-ray surface brightness inside and surrounding the cavities (equations 1 and 2 in \citealt{2021ApJ...923L..25U}), we find that the inner depressions represent 15\% deficits at a signal-to-noise ratio (SNR) of about 4.9, while the outer bubble represents a 7\% deficit at $\text{SNR} = 3.8$. {We do not detect an X-ray cavity at the position of the western outer lobe; this could be caused by a combination of projection effects and the influence of the external medium. The western outer lobe is less extended than the eastern one, which suggests it is more strongly projected along the line of sight. The western outer lobe also lies along the brighter part of the X-ray sloshing spiral, pointing to a contribution of bulk motions of the surrounding gas in bending the lobe towards the line of sight. Since X-ray cavity contrast decreases with increasing distance from the plane of the sky, projection effects may be hiding any X-ray cavity associated with the western outer lobe.}

In Fig. \ref{fig:chandra-condensation} we show the H$\alpha$-line intensity image (panel a) from the MUSE data alongside the {\it Chandra} unsharp masked image with H$\alpha$ and 340 MHz contours (panel {\em b}). As previously reported \citep{2010ApJ...721.1262M,2016MNRAS.460.1758H}, there is a nebula of warm gas at 10$^{4}$~K surrounding the cD galaxy of A496; the nebula is composed of a bright central region, coincident with the position of the radio core; several filaments of about 10~kpc in length, extended towards east; and a third, thin and faint filament extending towards NE. From the X-ray/H$\alpha$ comparison shown in panel (b), there is a significant spatial overlap between the filaments and the regions of enhanced X-ray emission, likely tracing denser (and cooler) gas. This is particularly evident along the X-ray bright ridge that extends eastward from the radio core. The warm gas is structured around the northern inner cavity/lobe, with the outermost filament circumventing the X-ray depression and heading further north. Due to the limitations in the FoV of the MUSE data (coincident with the FoV of Fig. \ref{fig:chandra-condensation}a), we are unable to tell if the filaments are even more extended. The average velocity dispersion of the filaments along the ridge is $\sigma_{v}=94\pm5$~km/s, whereas along the thin outer filament is $\sigma_{v}=65\pm5$~km/s. The decreasing velocity dispersion with increasing distance from the center may be consistent with the jets and possibly the ICM bulk motions being able to inject more kinetic energy in the central region (see \citealt{2019A&A...631A..22O}). However, a superposition of several unrelated structures with different velocities at the center of the nebula may also explain the higher central velocity dispersion.

\section{Discussion}
\subsection{Cycling of the central AGN}
Over the three visible outbursts in A496 (compact component, inner lobes, outer lobes), the NE-SW direction of propagation of the jets has remained nearly unchanged. At frequencies $\gax 1$ GHz, the central radio source contains a bright core, with possible inverted spectrum, and two-sided jets extending for about 4~pc in the NE-SW direction (as shown by \citealt{2024ApJ...961..134U}). On kpc scales, radio lobes are visible to the NE and SW of the core, with a steep radio spectrum and total synchrotron age of about 26~Myr (see Sect. \ref{subsec:radiolobes}). These lobes are associated with X-ray cavities (Fig. \ref{fig:chandra-condensation}), indicating that the jets have pushed aside the gas during their propagation. Two larger-scale fossil radio lobes emerge at lower frequency, located approximately along the same axis as the inner lobes, with shape slightly bent toward the same direction. Our analysis indicates that the outer lobes possess an ultra-steep radio spectrum, as typical of fading radio lobes produced in an earlier phase of activity. {\em Chandra} images show a large surface brightness
depression at the location of the NE outer lobe (Fig. \ref{fig:chandra-resid}{\em b}), suggesting that this region has likely been devoided of X-ray gas by the expanding radio bubble (see Sect. \ref{sec:connectionratioX}).
 
As noted in Sect. \ref{subsec:radiolobes}, we could only approximate the synchrotron ages of the outer lobes, finding a timescale of $\sim50 - 300$ Myr. An alternative method to estimate the age of the oldest outburst consists in deriving the age of the cavity associated with the NE outer lobe from the X-ray data. 
Given the diffuse, detached, and clearly buoyant morphology of the outer cavity, we derive its age using the buoyancy timescale (e.g., \citealt{2004ApJ...607..800B}):
\begin{equation}
\label{eq:tbuo}
    t_{b} = \frac{D_{cav}}{\sqrt{2gV/0.75S}}\,,
\end{equation}
where $D_{cav}$, $V$, and $S$ are the distance from the center, the volume, and the surface area of the cavity, respectively, while $g$ is the gravitational acceleration at the cavity position. To describe the geometry of the cavity (Fig.~\ref{fig:chandra-resid}), we considered a sphere with radius of 13~kpc at 33~kpc from the center (the bubble), from which we subtract another sphere with radius of 6~kpc at 26 kpc from the center (the wake). While this is a relatively simple approximation of the complex mushroom-wake morphology, we estimate that the impact on the final result (driven by $\sqrt{V/S}$ in Eq.~\ref{eq:tbuo}) is limited to $\leq$20\%, that we assume as our uncertainty. We measured the gravitational acceleration at the cavity position from the mass profile determined in \citet{2018ApJ...853..177P}, which accounts for the potential of the dark matter and of the cD galaxy, finding $g = 4.9\times10^{-8}$~cm~s$^{-2}$. 
\\Using Eq. \ref{eq:tbuo}, we find that the buoyant rise time of the cavity associated with the NE outer lobe is $t_{b} = 45\pm9$~Myr. This is longer than the total synchrotron age of the inner lobes ($26\pm$4~Myr), supporting the idea that the outer lobes are the oldest. Furthermore, projection effects may bias the $t_{b}$ estimated above towards shorter timescales than the real ones.

\subsection{Cooling of the ICM}\label{subsec:feedbackcooling}
\begin{figure}[ht!]
    \centering
    \includegraphics[width=0.95\linewidth]{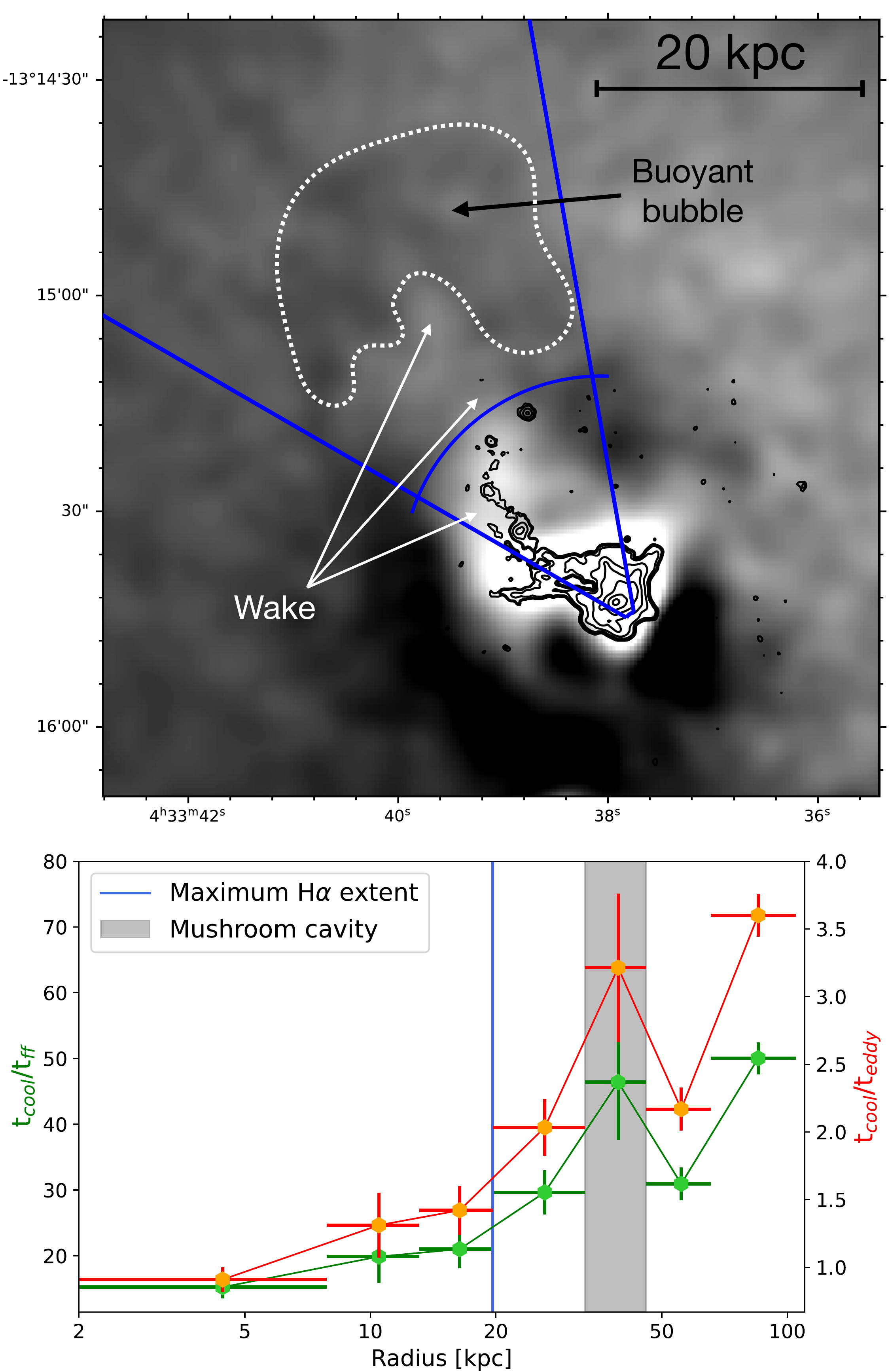}
    \caption{{\em Top}: {\it Chandra} unsharp masked image of A496, as shown in Fig. \ref{fig:chandra-resid}. Black contours show the morphology and extent of the warm gas nebula. The position of the wake of entrained gas is highlighted by arrows. The blue lines represent the sector used to extract the radial profiles of cooling instabilities shown in the {\em bottom} panel, and the blue arc shows the maximum extent of H$\alpha$ filaments. The dashed white region shows the mushroom head-shaped cavity. {\em Bottom}: Radial profiles of $t_{cool}/t_{ff}$ (in green) and of $t_{cool}/t_{eddy}$ (in orange) across the blue sector shown in the {\em top} panel. The blue vertical line marks the maximum extent of the detected H$\alpha$ filaments, while the gray-shaded area represents the radial extent of the outer X-ray cavity. Note that the profiles in the bottom panel are plotted further out than the extent of the blue sector in the top panel.} 
    \label{fig:chandra-new}
\end{figure}

Below the mushroom-shaped bubble, an elongated trail of X-ray bright gas is visible. Two scenarios can explain the formation of this trail in the X-ray map. Based on Fig. \ref{fig:chandra}\emph{b}, the X-ray bright gas extends from the cold front south of the central galaxy towards north-east. Thus, it is possible that the trail is an extension of this cold front, consisting of ICM that is oscillating in the potential of the cluster. Closer to the center, the ridge connecting the trail feature to the central galaxy may represent a stem of cool gas that is experiencing ram pressure from the hotter gas (see, for comparison, figure 7 in \citealt{2006ApJ...650..102A}). Similarly, the structure of the H$\alpha$ filaments (Fig. \ref{fig:chandra-new}\emph{b}) may have been generated by the same sloshing-induced motions. Alternatively, the X-ray trail and the corresponding H$\alpha$ filaments may be related to the rising mushroom-head cavity. Simulations of the uprising of X-ray cavities show that compression of the ICM entrained by the bubbles in their wake increases the cooling rate, thus generating colder material \citep{2007A&A...464..143G,2015ApJ...802..118B}. With time, the lower part of the trunk falls back to the cluster center, while the upper part remains at large radii. The shape of the X-ray wake in A496 is consistent with this picture, and the MUSE data provide further support. As visible in Fig. \ref{fig:chandra-new}, the warm gas filaments encompass the north inner bubble, and stretch towards the mushroom-shaped cavity. Thus, it is possible that at least part of the warm gas may be the end-product of ICM cooling in the wake of the buoyant bubble. 

\par Different observational strategies have been proposed to verify if the cooling of the hot gas is efficient enough to stimulate the condensation into warm gas filaments. A ratio of $\leq20-30$ between the cooling time $t_{cool}$ and the free fall time $t_{ff}$ (e.g., \citealt{2015ApJ...799L...1V}) has been proposed as a sensitive predictor of the radial range where cooling is expected to occur. 
Alternatively, \citet{2018ApJ...854..167G} proposed that cooling instabilities are expected to develop when the ratio between the cooling time and the eddy turnover time, $t_{\text{eddy}}$ (that is the time at which a turbulent vortex gyrates and produces density fluctuations), is $\lessapprox$1. We tested these ratio in A496 by deriving radial profiles of the above quantities ($t_{cool}$, $t_{ff}$, $t_{eddy}$) in a narrow wedge ($50^{\circ}$-wide) centered on the radio core and encompassing the X-ray bright wake (see Fig. \ref{fig:chandra-new}). The radial bins have been chosen to collect at least 4000 net counts in the 0.5 - 7 keV band per bin. We derived the:
\begin{itemize}[noitemsep, leftmargin=*]
    \item Cooling time profile. By fitting a \texttt{projct$\ast$tbabs$\ast$apec} model using \texttt{XSPEC} to the spectra of the radial bins, we derived the deprojected temperature $kT$ and electron density $n_e$ of the ICM. Then, we measured the cooling time as:
    \begin{equation}
    t_{\text{cool}} = \frac{\gamma}{\gamma -1} \frac{kT}{\mu \,X \,n_{\text{e}}\,\Lambda(\text{T, Z})}\,\text{,}
    \label{tcool}
\end{equation}
where $\gamma = 5/3$ is the adiabatic index, $\mu\approx0.6$ is the mean molecular weight, $X\approx0.7$ is the hydrogen mass fraction and $\Lambda(\text{T,Z})$ is the cooling function (from \citealt{1993ApJS...88..253S}). We find that the cooling time has its minimum of $t_{cool} = 370\pm23$~Myr in the innermost bin ($r\leq8$~kpc). \\
    \item Free fall time profile. The free fall time is defined as:
\begin{equation}
    \label{tfreefall}
        t_{ff}(r) = \sqrt{\frac{2r}{g(r)}}  
\end{equation}
\noindent where $g(r)$ is the gravitational acceleration at distance $r$. We measured $g(r)$ from the mass profile determined in \citet{2018ApJ...853..177P}. \\
    \item Eddy turnover time profile. The eddy turnover time is defined as:
\begin{equation}
    \label{teddy}
    t_{\text{eddy}} = 2\pi\,\frac{r^{2/3}\,L^{1/3}}{\sigma_{v,3D}}
\end{equation}
\noindent where $L$ is the injection scale of turbulence (usually assumed to be described by the extent of the warm gas filaments), and $\sigma_{v,3D}$ is the 3D velocity dispersion of the gas \citep{2018ApJ...854..167G}. We measured the $t_{eddy}$ profile in the concentric wedges assuming $L = 20$~kpc (the length of the H$\alpha$-filaments, see Fig. \ref{fig:chandra-condensation}), and using $\sigma_{v,3D} =\sqrt{3}\times \langle\sigma_{v}\rangle = 133\pm9$~km/s measured from MUSE data.  
\end{itemize}
We plot in Fig. \ref{fig:chandra-new} (bottom) the resulting radial profiles of $t_{cool}/t_{ff}$ and $t_{cool}/t_{eddy}$. Both cooling thresholds are met up to the maximum extent of the filaments, as similarly found in studies of larger samples of clusters and groups (e.g., \citealt{2019A&A...631A..22O}). The profiles steeply rise outwards, indicating that cooling becomes less efficient with increasing radius. However, we note that the fourth radial bin, at $20\leq r \leq33$~kpc, shows $t_{cool}/t_{ff} = 29$ and $t_{cool}/t_{eddy} = 2$, close to the cooling thresholds. Thus, warm gas might be trailing the mushroom head-shaped cavity along the full length of the wake.

Only future optical IFU observations with a larger FoV (e.g., with SITELLE, \citealt{2012SPIE.8446E..0UG}) or with an offset pointing (e.g., with MUSE) can reveal any faint warm gas filament extending all the way to the tip of the wake behind the mushroom head-shaped cavity. Such configuration has been revealed so far only in the case of the Perseus cluster. In that case, the warm gas filaments behind the NW buoyant bubble have a clear horseshoe morphology, since the filaments bend in opposite directions either side of the center of the bubble. The horseshoe filaments may be tracing the toroidal flow pattern induced by the NW bubble of Perseus in its wake (e.g., \citealt{2003MNRAS.344L..48F,2006MNRAS.367..433H,2024ApJ...962...96V}). Collecting similar morphological and kinematical information for A496 will provide further constraints to cooling instabilities, and will reveal whether the cooling of the ICM into warm gas has been stimulated by the uprise of the mushroom-head cavity or by the sloshing motion.

\subsection{Non-detection of a mini-halo in A496}

In several cool core galaxy clusters it is possible to observe diffuse radio sources, referred to as {\em mini-halos}, that extend over the whole cool core region. Radio mini-halos are faint, diffuse, and often associated with sloshing clusters (e.g., \citealt{2013ApJ...762...78Z,2018A&A...617A..11G,2019ApJ...880...70G,2024A&A...686A..82B}). Moreover, the boundary of mini-halos seem to align with cold fronts in the ICM, suggesting that the turbulence generated by the sloshing might energize the relativistic particles traced by their synchrotron emission. Despite ongoing vigorous gas sloshing, and the presence of a favorable source of seed electrons (the central radio galaxy), current radio observations fail to detect a mini-halo in A496. It is likely that the sensitivity and $uv$ coverage of the available images are inadequate to detect very faint diffuse radio emission. 

\citet{2019ApJ...880...70G} provided a relation between the 1.4 GHz luminosity of mini-halos and the X-ray luminosity of the ICM within the central 70 kpc. Based on such relation, we can estimate the expected level of diffuse radio emission, if A496 were to contain a mini-halo. 
From the {\em Chandra} data we measure an X-ray bolometric luminosity of $1.2\times 10^{44}$~erg~s$^{-1}$ for the inner 70 kpc of A496. 
From the best-fit relations of \citet{2019ApJ...880...70G}, we estimate a 1.4 GHz radio power of about $2\times10^{22} - 4 \times10^{22}$ W Hz$^{-1}$, which translates into a flux density of $\sim 10-20$ mJy. For comparison, using the relation of \citet{2020MNRAS.499.2934R} between the 1.4 GHz mini-halo power and the X-ray luminosity within 600 kpc we would expect a radio power of about $10^{23}$ W Hz$^{-1}$. Thus, the relation of \citet{2019ApJ...880...70G} provides the most stringent limit. If the above putative flux of $\sim 10-20$ mJy was uniformly spread over the innermost 70 kpc (that is also the distance of the northern cold front from the center), we would expect an average surface brightness of $\sim 0.3-0.6$ $\mu$Jy/arcsec$^2$ at 1.4 GHz, of $\sim 0.7-1.4$ $\mu$Jy/arcsec$^2$ at 617 MHz and of $\sim 1.3-2.6$ $\mu$Jy/arcsec$^2$ at 330 MHz (assuming a spectral index $\alpha=1$). These estimates demonstrate the current non-detectability of a mini-halo, since such faint emission is well below the sensitivity of the available radio images (4 $\mu$Jy/arcsec$^2$ at 330 MHz, 12 $\mu$Jy/arcsec$^2$ at 617 MHz).

\section{Summary}

In this work, we presented deep, multi-frequency radio images of the cool core cluster A496, that allowed us to detect three distinct jet episodes. On sub-kpc scales there is evidence of an ongoing SMBH activity with a possibly inverted radio spectrum ($\alpha\sim0.7$ above 5 GHz and $\alpha\sim0$ below). On scales of $\sim20$~kpc, an older activity inflated radio lobes with a steep spectral index ($\alpha=2.0\pm0.1$). Our modeling of their synchrotron spectrum indicates that the jet activity lasted for about $\sim20$~Myr, before switching off about 6~Myr ago. On larger scales ($\sim50-100$~kpc), the GMRT images reveal an oldest episode of activity with an ultra-steep spectrum ($\alpha=2.7\pm0.2$). 

The archival {\it Chandra} data revealed X-ray cavities corresponding to the lobes of the older activity, and to the north-eastern lobe of the oldest activity. The outer cavity has the shape of a mushroom-head, typical of late-time buoyant motions of bubbles. Based on the buoyancy timescale, we estimate a travel time of $\sim45$~Myr. Furthermore, there is an X-ray bright feature trailing the fossil lobe/cavity, along which the hot gas has efficiently cooled. Indeed, warm gas filaments traced by their H$\alpha$ emission are seen stretching towards the north-eastern outer bubble. Condensation of the ICM may have been stimulated by the rising cavity in its wake, or by uplift due to sloshing motions. The first scenario would be supported if future optical IFU observations with an offset pointing or with a larger field of view reveal that warm filaments extend to the tip of the wake.

Despite the vigorous sloshing occurring in A496, our radio images fail to detect a radio mini-halo within the cool core. Based on scaling relations between the radio power of mini-halos and the X-ray luminosity of the host cool core, we estimate that a sensitivity between $\sim 0.3-3$ $\mu$Jy/arcsec$^2$ (depending on frequency), which is well below that of the radio images presented in this work, would be required to detect the faint, diffuse emission. Future observations with the upgraded GMRT, the JVLA, or the Square Kilometer Array precursors ASKAP and MeerKAT may provide the sensitivity and $uv$ coverage needed to unveil an extremely faint mini-halo in this sloshing cool core, further supporting the connection between sloshing and mini-halo formation, and extending it down to smaller masses of the host cluster.

\begin{acknowledgements}
{We thank the anonymous reviewer for their useful comments to our manuscript}. Basic research in radio astronomy at the Naval Research Laboratory is supported by 6.1 Base funding. Construction and installation of VLITE was supported by the NRL Sustainment Restoration and Maintenance fund. The National Radio Astronomy Observatory is a facility of the National Science Foundation operated under cooperative agreement by Associated Universities, Inc. The scientific results reported in this article are based on data obtained from the GMRT Data Archive. We thank the staff of the GMRT that made the observations possible. GMRT is run by the National Centre for Astrophysics of the Tata Institute of Fundamental Research. This scientific work makes use of the data from the Murchison Radio-astronomy Observatory and Australian SKA Pathfinder, managed by CSIRO. Support for the operation of the MWA and ASKAP is provided by the Australian Government (NCRIS). ASKAP and MWA use the resources of the Pawsey Supercomputing Centre. Establishment of ASKAP, MWA and the Pawsey Supercomputing Centre are initiatives of the Australian Government, with support from the Government of Western Australia and the Science and Industry Endowment Fund. We acknowledge the Wajarri Yamatji people as the traditional owners of the observatory sites. This paper employs a list of Chandra datasets, obtained by the Chandra X-ray Observatory, contained in the Chandra Data Collection (CDC)~\url{https://doi.org/10.25574/cdc.241}. This research has made use of software provided by the Chandra X-ray Center (CXC) in the application packages CIAO. The results are based on observations collected at the European Southern Observatory under ESO programme 094.B-059.

\end{acknowledgements}

%
%

\end{document}